\documentclass[
10pt,
aps,
notitlepage, 
bibnotes,
prb,
twocolumn,
showpacs,preprintnumbers,superscriptaddress,
amsmath,amssymb,floatfix]{revtex4-2}

\usepackage{graphicx}
\usepackage{dsfont}
\usepackage{dcolumn}
\usepackage{bm}
\usepackage{color}
\usepackage{xcolor}
\usepackage{url}
\usepackage{tabularx}
\usepackage{array}
\usepackage{booktabs}
\usepackage{multirow}
\usepackage{comment}
\usepackage[colorlinks,citecolor=red,linkcolor=blue,urlcolor=blue]{hyperref}
\usepackage{footnote}
\usepackage{lipsum}
\usepackage[normalem]{ulem}
\usepackage{booktabs}

\usepackage{csquotes}
\usepackage{listings}
\definecolor{colorA}{rgb}{0, 0, 1}
\definecolor{colorB}{rgb}{0.5, 0, 0.9}
\definecolor{colorC}{rgb}{0.4, 0, 0.4}
\definecolor{color_green}{rgb}{0, 0.39, 0}

\makeatletter
  \def\my@tag@font{\normalsize}
  \def\maketag@@@#1{\hbox{\m@th\normalfont\my@tag@font#1}}
  \let\amsmath@eqref\eqref
  \renewcommand\eqref[1]{{\let\my@tag@font\relax\amsmath@eqref{#1}}}
\makeatother

\begin{document}

\def\afflux{Department of Physics and Materials Science, University of Luxembourg, L-1511 Luxembourg, Luxembourg}

\title{Regularized Micromagnetic Theory for Bloch Points}

\author{Vladyslav~M.~Kuchkin}
\email{vladyslav.kuchkin@uni.lu}
\affiliation{\afflux}

\author{Andreas Haller}
\affiliation{\afflux}

\author{Andreas Michels}
\affiliation{\afflux}

\author{Thomas L. Schmidt}
\affiliation{\afflux}

\author{Nikolai~S.~Kiselev}
\affiliation{Peter Gr\"unberg Institute, Forschungszentrum J\"ulich and JARA, 52425 J\"ulich, Germany}

\date{\today}

\maketitle

\textbf{
Magnetic singularities known as Bloch points (BPs) present a fundamental challenge for micromagnetic theory, which is based on the assumption of a fixed magnetization vector length.
Due to the divergence of the effective field at a BP, classical micromagnetics fails to adequately describe BP dynamics.
To address this issue, we propose a regularized micromagnetic model in which the magnetization vector can vary in length but not exceed a threshold value.
More specifically, the magnetization is treated as an order parameter constrained to a $\mathbb{S}^3$-sphere.
This constraint respects fundamental properties of local spin expectation values in quantum systems.
We derive the corresponding regularized Landau–Lifshitz–Gilbert equation and the analogue of the Thiele equation describing the steady motion of spin textures under various external stimuli.
We demonstrate the applicability of our theory by modeling the dynamics of several magnetic textures containing BPs, including domain walls in nanowires, chiral bobbers, and magnetic dipolar strings.
The presented results extend micromagnetic theory by incorporating a regularized description of BP dynamics.
}

\noindent
Micromagnetism~\cite{brown1963micromagnetics,aharoni2000introduction, chikazumi1997physics,hubert2008magnetic} is a well-established classical field theory that describes both static and dynamic properties of magnetic media, including bulk and nanoscale crystals, amorphous alloys, and heterostructures.
The success of micromagnetism lies in its strong predictive power.
It accurately describes, for example, the dynamics of domain walls, resonance spectra, magnetization reversal loops, and many other experimentally observed phenomena~\cite{hubert2008magnetic}.
Micromagnetism has profoundly influenced the development of technologies based on magnetic materials, including computing devices~\cite{Chumak2014, Parkin2008, Wolf2001}, data storage~\cite{ODell_86,Malozemoff_79,Victora2005}, electric motors~\cite{Coey2002}, and robotics~\cite{Huang2016, Wu2020}.

In micromagnetism, the magnetization in a ferromagnet is considered a continuous three-dimensional vector field $\mathbf{n}(\mathbf{r})$, defined at every point $\mathbf{r}$ within the sample.
Continuity of the magnetization field is a central assumption of micromagnetic theory—and, at the same time, one of its key limitations.
It holds for a broad but nevertheless limited class of magnetic spin textures.
In particular, only a few years after Brown formulated the foundations of micromagnetism\cite{brown1963micromagnetics}, Feldtkeller~\cite{Feldtkeller1965} and Döring~\cite{Doring1968} independently demonstrated that, in certain configurations representing statically stable solutions of the micromagnetic Hamiltonian, this continuity is broken.
The solutions they identified correspond to hedgehog-like vector fields, $\mathbf{n} = \mathbf{r}/r$, stabilized by boundary conditions in ball-shaped samples.
Today, such point-like topological defects are commonly referred to as Bloch points (BPs).

Subsequent studies have shown that BPs can be stabilized under various conditions.
For instance, BPs arise naturally during dynamic processes, such as the nucleation and annihilation of topologically non-trivial spin textures, including merons~\cite{Rybakov_25}, skyrmions~\cite{Bogdanov1989, Tokura2020}, and hopfions~\cite{Voinescu2020, Rybakov2022, Zheng_23}.
Therefore, a consistent theory that allows for the modeling of BP dynamics is essential for constructing a unified physical picture of topological magnetic solitons.

The existence of BPs and their mobility in the magnetic samples are supported by numerous direct and indirect experimental observations~\cite{Slonczewski1975, Slonczewski1981, Slonczewski1976, Bullock1974, Hasegawa1975, Josephs1976, Milde2013, Fruchart_14, Jamet2015, Im_19, zheng_18bobber, Im_19,wartelle_19, Gomez_25, DaCol2014, Wohlhter2015, Kanazawa2016, Donnelly2017, Donnelly2020, Han2021, Schbitz2021, Arekapudi2021, Cocoons_22, Savchenko2022, Yasin2024, Han2024, Kern2025}, 
and has a long history of theoretical study~\cite{Hubert1975, Arrott1979, Aharoni1980, Sokalski1980, kufaev1988vibrations, kufaev1989dynamics, kabanov1989bloch, kufaev1992effects, Galkina1993,  Thiaville1994,   Forster2002, Thiaville2003, VanWaeyenberge2006, Komineas2007, Volkov2008, Elas2011, Lebecki2012, Pylypovskyi2012, Zverev2013, Schtte2014, Andreas2014, CarvalhoSantos2015, Pylypovskyi2015, DeLucia2017, Ackerman2017, Beg2019, Charilaou2020, Mller2020, Gorobets2020, Pathak2021, DeRiz2021, Birch2021, Azhar2022, Sez2022, Kuchkin2023, ZambranoRabanal2023, CastilloSeplveda2023, Lang2023, Qiu2024, HermosaMuoz2024, Kim2024, Shimizu2025, Yastremsky2025, Winkler2025}.
In contrast, describing BP dynamics remains a fundamental challenge for micromagnetic theory.
The obstacle lies in the divergence of the effective field in the magnetization dynamics equation when a BP is present.
This divergence stems directly from the constraint of constant magnetization magnitude.
While this constraint in some circumstances is consistent with the results of quantum spin models and holds for smooth textures, recent studies have shown that quantum fluctuations remain significant in the vicinity of BPs~\cite{Elas2014, Tapia2024, Kuchkin2025}.
These fluctuations can substantially reduce the length of observable classical spins, but cannot increase it.
Thus, the magnetization must satisfy the inequality $|\mathbf{n}(\mathbf{r})| \leq 1$, rather than being strictly constrained to unit length.
Such a relaxed constraint enables the use of an order parameter that remains regular even in the presence of Bloch points, thereby resolving the associated divergence.
This motivated us to develop a regularized micromagnetic theory capable of describing textures both with and without BPs.
Given the proven predictive power of classical micromagnetics, such a theory should reproduce all well-established results and modify only those solutions that contain BPs.
Here, we present such a regularized micromagnetic theory.

The paper is organized as follows. 
We first formulate the problem and introduce the new order parameter along with the regularized micromagnetic Hamiltonian, which we refer to as the $\mathbb{S}^3$-model.
We then derive the dynamical equation for the $\mathbb{S}^3$-model, including the effects of external torques.
Next, using the collective coordinate approach, we derive an analog of the Thiele equation describing the rigid motion of magnetic textures within this framework.
The applicability of the derived equations is illustrated through examples of the dynamics of various BP-hosting spin textures.
We also compare numerical simulations based on the $\mathbb{S}^3$-model and the standard micromagnetic model ($\mathbb{S}^2$-model) with analytical solutions of the Thiele equations.
Finally, we discuss the topological properties of the $\mathbb{S}^3$-model, review earlier experiments on the dynamics of magnetic bubbles containing BPs, and address several frequently asked questions concerning our approach, its numerical implementation, and previously compare it with earlier attempts to regularize micromagnetic equations.
Additional details on the model parameters used in the micromagnetic simulations are provided in the Methods section.

\vspace{0.5cm}
\noindent
\textbf{Results}

\noindent
In this study, we consider the following form of the micromagnetic Hamiltonian: 
\begin{equation}
    E=\intop\left[e_\mathrm{exi} + e_\mathrm{dmi} + e_\mathrm{ddi} + e_\mathrm{ani} + e_\mathrm{z}\right]\mathrm{d}V,\label{micro_hamiltonian}
\end{equation}
where $e_\mathrm{exi}=\mathcal{A}\left(\nabla\mathbf{n}\right)^{2}$ is the Heisenberg exchange interaction, $e_\mathrm{dmi}=\mathcal{D}\mathbf{n}\cdot\nabla \times\mathbf{n}$ is the Dzyaloshinskii-Moriya interaction (DMI), $e_\mathrm{ani}=-\mathbf{n}\cdot \hat{\mathcal{K}}\cdot \mathbf{n}$ is the magnetic anisotropy, $e_\mathrm{ddi}=-\frac{1}{2}\mathcal{M}\mathbf{B}_\mathrm{d}\cdot\mathbf{n}$ is the demagnetizing field (or dipole-dipole) interaction, and $e_\mathrm{z}=-\mathcal{M}\mathbf{B}_\mathrm{ext}\cdot\mathbf{n}$ is the Zeeman interaction with the external magnetic field. 
We denote the saturation magnetization by $\mathcal{M}$, and define the normalized magnetization as $\mathbf{n} = \mathbf{M}(r)/\mathcal{M}$.
The Hamiltonian~\eqref{micro_hamiltonian} can be straightforwardly extended to include additional interactions.
We adopt this form of the Hamiltonian because it covers all example cases considered in this study.

Interestingly, the equation describing magnetization dynamics was established by Landau and Lifshitz~\cite{LandauLifshitz1935} even before micromagnetic theory itself was formulated.
Today, this equation is commonly referred to as the Landau–Lifshitz–Gilbert (LLG) equation. Its standard form, expressed in terms of the unit magnetization vector, is given by:
\begin{equation}
\dot{\mathbf{n}} = -\gamma \mathbf{n} \times \mathbf{b} - \alpha \gamma \mathbf{n} \times (\mathbf{n} \times \mathbf{b}), \label{LLG_micro}
\end{equation}
where $\dot{\mathbf{n}}$ denotes the time derivative of the magnetization vector, $\gamma = \gamma_0 / (1 + \alpha^2)$, $\gamma_0$ is the electron gyromagnetic ratio, and $\alpha$ is the Gilbert damping constant.
The effective field is defined as $\mathbf{b} = -\mathcal{M}^{-1} \delta E / \delta \mathbf{n}$, where $E$ is the micromagnetic Hamiltonian ~\eqref{micro_hamiltonian}.
By definition, Eq.~\eqref{LLG_micro} preserves the magnitude of the magnetization vector.
Applying it to a magnetic texture containing a BP inevitably leads to a divergence problem, which can be illustrated by considering only the exchange energy term in Eq.~\eqref{micro_hamiltonian}.
The effective field due to exchange interaction is
$\mathbf{b}_\mathrm{exi} = -2\mathcal{A} \mathcal{M}^{-1} \nabla^2 \mathbf{n}$.
For hedgehog-like spin textures, it diverges $|\mathbf{b}_\mathrm{exi}| \sim 1/r^2$ as $r \to 0$.
The only way to resolve the divergence problem is to allow the absolute value of magnetization to vary continuously, reaching $|\mathbf{n}| = 0$ at $r = 0$, i.e., at the core of the BP (see Supplementary material~I).

However, in this case, the order parameter no longer lies on the two-sphere $\mathbb{S}^2$, and one must introduce appropriate corrections to both the model Hamiltonian~\eqref{micro_hamiltonian} and the LLG equation~\eqref{LLG_micro}.
The key question is on which manifold the new order parameter should be defined.

Our previous study of BPs in the quantum spin model~\cite{Kuchkin2025} showed that an effective micromagnetic model can be formulated in terms of a magnetization field defined on the three-sphere $\mathbb{S}^3$.
In other words, the order parameter can be defined as a four-dimensional vector $\bm{\nu} = (\nu_1, \nu_2, \nu_3, \nu_4)$ constrained to $\mathbb{S}^3$, i.e., $|\bm{\nu}| = 1$ at every point in the magnetic sample.
The first three components of $\bm{\nu}$ correspond to the Cartesian components of the magnetization vector $\mathbf{n}$,
while the fourth component encodes the magnetization length via the relation
\begin{equation}
\nu_1 = n_x, \quad \nu_2 = n_y, \quad \nu_3 = n_z, \quad \nu_4^2 = 1 - |\mathbf{n}|^2.
\label{nu_m}
\end{equation}
As follows from \eqref{nu_m}, for $|\bm{\nu}|=1$, magnetization satisfies the inequality $|\mathbf{n}|\leq1$, and thus magnetization can be reduced up to zero but never exceeds the maximal value ($|\mathbf{M}(\mathbf{r})| \leq \mathcal{M}$). 

To rewrite the Hamiltonian~\eqref{micro_hamiltonian} in terms of the new order parameter, $E(\mathbf{n}) \mapsto \mathcal{E}(\bm{\nu})$, we follow the approach of Ref.~\cite{Kuchkin2025} and generalize only the Heisenberg exchange interaction, $e_\mathrm{exi}(\mathbf{n}) \mapsto e_\mathrm{exi}(\bm{\nu})$, as follows:
\begin{equation}
e_\mathrm{exi}(\bm{\nu}) = \mathcal{A} \sum_{i=1}^{4} (\nabla \nu_i)^2 + \kappa \nu_4^2.
\label{exchange_nu}
\end{equation}
All other terms in Eq.\eqref{micro_hamiltonian} are obtained by a straightforward substitution $(n_x, n_y, n_z) \mapsto (\nu_1, \nu_2, \nu_3)$.

The phenomenological parameter $\kappa$ in Eq.~\eqref{exchange_nu} is assumed to be positive and has units of J/m$^3$.
Like all other material parameters, it is treated as a temperature-dependent quantity, $\kappa=\kappa(T)$.
This parameter also serves to distinguish the \textit{fictitious} component $\nu_4$ from the physically measurable components of the order parameter, $\nu_1$, $\nu_2$, and $\nu_3$.
We refer to $\nu_4$ as fictitious because it has no direct physical observable associated with it and is introduced solely to extend the order parameter space from $\mathbb{S}^2$ to $\mathbb{S}^3$ for regularization purposes.

The regularized form of the exchange energy term~\eqref{exchange_nu} ensures the continuity of the order parameter, even for hedgehog-like configurations.
In such cases, the measurable components of the order parameter can vanish at the center of the texture and approach unity exponentially, as $\sim 1 - \exp(-r / L_\mathrm{\kappa})$, far from the Bloch point~\cite{Kuchkin2025}.
The parameter $L_\mathrm{\kappa} = \sqrt{\mathcal{A} / \kappa}$, which has units of length, can be interpreted as the characteristic size of the BP.

The effective field $\bm{\beta}(\bm{\nu}) = -\mathcal{M}^{-1} ( \delta \mathcal{E} / \delta \bm{\nu})$, acting on the four-dimensional vector $\bm{\nu}$, can be computed directly from the regularized Hamiltonian $\mathcal{E}(\bm{\nu})$.
Together with $\bm{\nu}(\mathbf{r})$, the effective field $\bm{\beta}(\bm{\nu})$ remains continuous throughout the sample, even in the presence of BPs.
In the next section, using the above expression for effective field and following the approach of Landau and Lifshitz~\cite{LandauLifshitz1935}, we derive a generalized dynamical equation for the new order parameter $\bm{\nu}$.

\begin{figure}[tb!]
\centering
\includegraphics[width=8.0cm]{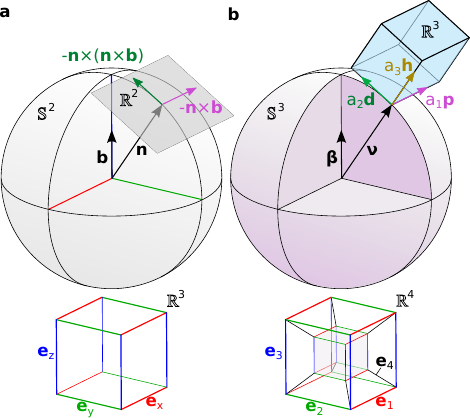}
\caption{~\small \textbf{Geometric interpretation of the LLG equations in the $\mathbb{S}^2$ and $\mathbb{S}^3$ models.} 
Panels \textbf{a} and \textbf{b} illustrate the configuration spaces of the order parameter for the standard micromagnetic model constrained to the two-sphere $\mathbb{S}^2$, and the regularized model defined on the three-sphere $\mathbb{S}^3$, respectively.
In \textbf{a}, the three-dimensional unit vector $\mathbf{n}$ lies on the two-sphere embedded in $\mathbb{R}^3$ space. The associated tangent space $\mathbb{R}^2$ contains orthogonal directions corresponding to the conventional basis vectors: $\mathbf{n} \times \mathbf{b}$ (precession) and $\mathbf{n} \times (\mathbf{n} \times \mathbf{b})$ (damping).
In \textbf{b}, the four-dimensional unit vector $\bm{\nu}$ is defined on the three-sphere $\mathbb{S}^3$ embedded in $\mathbb{R}^4$ space.
Note that in panel \textbf{b}, the vectors $\bm{\nu}$, $\mathbf{p}$, $\mathbf{d}$, and $\mathbf{h}$ are mutually orthogonal in four-dimensional space; this orthogonality is represented schematically, as it cannot be faithfully depicted in a two-dimensional figure.
} 
\label{Fig0}
\end{figure}

\vspace{0.5cm}
\noindent
\textbf{Regularized dynamics equation}.
\noindent
The LLG equation~\eqref{LLG_micro} has a simple geometric interpretation: it describes the motion of a point constrained to the surface of a two-sphere $\mathbb{S}^2$ [Fig.~\ref{Fig0}(a)].
In this case, $\dot{\mathbf{n}}$ is always perpendicular to $\mathbf{n}$ and lies in the tangent plane to the sphere, which is locally isomorphic to $\mathbb{R}^2$.
Within this plane, $\dot{\mathbf{n}}$ is spanned by two orthogonal basis vectors: $-\mathbf{n} \times \mathbf{b}$ and $-\mathbf{n} \times (\mathbf{n} \times \mathbf{b})$, which are usually called \textit{precession} and \textit{dissipation} term, respectively.
These vectors form a complete basis of the tangent space, and $\dot{\mathbf{n}}$ can be written as a linear combination of basis vectors (cf. Eq.~\eqref{LLG_micro}).

\begin{figure*}[tb!]
\centering
\includegraphics[width=18cm]{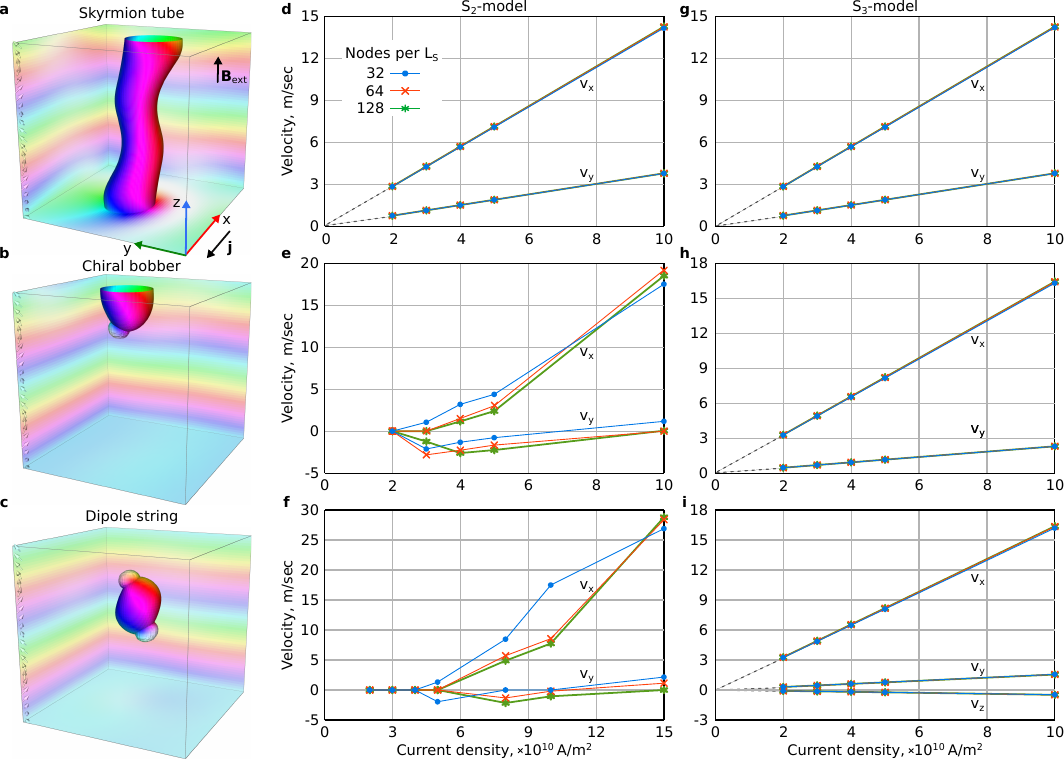}
\caption{~\small \textbf{Current-induced dynamics of various magnetic solitons in chiral magnets}. 
% %
\textbf{a}, \textbf{b} and \textbf{c} show skyrmion tube, chiral bobber, and dipolar string, respectively.
Corresponding magnetic texture are visualized by color-coded magnetization at the isosurfaces ($\nu_{3}=0$) and edges of the simulated domain.
Additional isosurfaces having nearly spherical shapes indicate the position of BPs in (b) and (c) is given by $\nu_{4}=0.9$.
The first column of plots in \textbf{d}, \textbf{e}, and \textbf{f} show velocity dependencies on the current density for skyrmion, chiral bobber, and dipole string, respectively, estimated from micromagnetic simulations based on the standard $\mathbb{S}^{2}$-model. 
The second column of plots in \textbf{g}, \textbf{h}, and \textbf{i} shows velocity dependence estimated from micromagnetic simulations based on the regularized $\mathbb{S}^{2}$-model. 
Every plot contains velocities estimated in the simulations performed with different discretizations, see legend in \textbf{d}.
For details, see the main text and Method.
} 
\label{Fig1}
\end{figure*}

In the case of dynamics on the $\mathbb{S}^{3}$-sphere [Fig.~\ref{Fig1}(b)], the corresponding vector $\dot{\bm{\nu}}$ is perpendicular to $\bm{\nu}$ and lays in the tangent space of $\mathbb{S}^{3}$, which is locally isomorphic to $\mathbb{R}^{3}$.
Within this $\mathbb{R}^{3}$ space, the vector $\dot{\bm{\nu}}$ can be spanned by three basis vectors, $\mathbf{p}$, $\mathbf{d}$ and $\mathbf{h}$, and can be written as linear combination:
\begin{equation}
\dot{\bm{\nu}}=a_{1}\mathbf{p}+a_{2}\mathbf{d}+a_{3}\mathbf{h},
\label{RLLG_ini}
\end{equation}
where $a_{1}$, $a_{2}$, $a_{3}$ are scalars that generally speaking can be functions of $\bm{\nu}$. 
These basis vectors are required, by definition, to be mutually orthogonal and orthogonal to $\bm{\nu}$.
Without loss of generality, we define the first basis vector $\mathbf{p}$ as a natural generalization of the precession term in Eq.~\eqref{LLG_micro}, meaning it is chosen to be orthogonal to both $\bm{\nu}$ and $\bm{\beta}$.
In four dimensions, however, this condition alone does not uniquely determine $\mathbf{p}$ -- a third orthogonal direction must also be fixed.
To resolve this ambiguity, we arbitrarily assume that $\mathbf{p}$ is orthogonal to the basis vector $\mathbf{e}_4$:
\begin{equation}
    \mathbf{p} =\bm{\nu}\times\bm{\beta}\times\mathbf{e}_{4}.
    \label{p_term}
\end{equation}
The right-hand side of \eqref{p_term} denotes the Hodge dual of the wedge product, which is equivalent to computing the determinant of a $4 \times 4$ matrix whose upper row consists of the standard $\mathbb{R}^{4}$ basis vectors $(\mathbf{e}_1, \mathbf{e}_2, \mathbf{e}_3, \mathbf{e}_4)$ and the next three rows contain components of the four-dimensional vectors $\bm{\nu}$, $\bm{\beta}$, and $\mathbf{e}_4=(0,0,0,1)$.

The second basis vector $\mathbf{d}$ is defined by analogy with the damping term in Eq.~\eqref{LLG_micro}:
\begin{equation}
\mathbf{d} = \bm{\nu} \left(\bm{\nu} \cdot \bm{\beta} \right) - \bm{\beta}.
\label{d_term}
\end{equation}
Since $\mathbf{d}$ is a linear combination of $\bm{\nu}$ and $\bm{\beta}$, both of which are orthogonal to $\mathbf{p}$, it follows that $\mathbf{d}$ is also orthogonal to $\mathbf{p}$.
Finally, the third basis vector $\mathbf{h}$ is defined as orthogonal to $\bm{\nu}$, $\mathbf{p}$ and $\mathbf{d}$:
\begin{equation}
\mathbf{h} = \bm{\nu}\times\mathbf{p}\times\mathbf{d},
\label{f_term}
\end{equation}
where, similar to \eqref{p_term}, the right-hand side denotes the Hodge dual of the wedge product.

Our goal is to derive dynamical equations that reduce to the standard LLG equation~\eqref{LLG_micro} in the absence of BPs.
By setting $\nu_4 = 0$ in Eqs.~\eqref{p_term}–\eqref{f_term}, we find that Eq.~\eqref{RLLG_ini} recovers the standard LLG form when the prefactors are chosen as $a_1 = -\gamma$ and $a_2 = -\alpha\gamma$.
The coefficient $a_3$ cannot be determined in this limiting case, as the vector $\mathbf{h}$ is quadratic in the effective field and has no counterpart in the standard LLG equation~\eqref{LLG_micro}.
To determine $a_1$ and $a_2$ in this limit, one must formally set $a_3$ to zero.
For generality, we define $a_3 = \epsilon\gamma$, where $\epsilon$ is a new phenomenological constant with units of 1/Tesla.
Taking all of the above into account, the regularized LLG equation takes the form:
\begin{equation}
\dot{\bm{\nu}} = -\gamma \mathbf{p} - \alpha\gamma \mathbf{d} - \epsilon\gamma\, \bm{\nu} \times \mathbf{p} \times \mathbf{d}. \label{RLLG}
\end{equation}
In the following, we demonstrate that, in the first approximation, the last term in Eq.~\eqref{RLLG} can be omitted.
While the first two terms are linear in $|\bm{\beta}|$, the third one is quadratic.
In Supplementary Material~II, we prove that it is impossible to construct $\mathbf{h}$ linear in components of $\bm{\beta}$.
The fact that $|\mathbf{h}|\sim\mathcal{O}\left(\bm{\beta}^2\right)$ suggests that it might dominate at large $\bm{\beta}$.
In practice, however, it is not the case.
First, note that all three terms in Eq.~\eqref{RLLG} depend only on the component of the effective field orthogonal to $\bm{\nu}$, denoted $\bm{\beta}_\perp$.
The longitudinal component, $\bm{\beta}_\parallel=(\bm{\nu} \cdot \bm{\beta}) \bm{\nu}$, does not contribute to $\dot{\bm{\nu}}$, and we can replace $\bm\beta$ in Eqs.\eqref{p_term}–\eqref{f_term} with $\bm{\beta}_\perp = \bm{\beta} - (\bm{\nu} \cdot \bm{\beta}) \bm{\nu}$.
It can be seen that the first two terms scale linearly with $\bm{\beta}_\perp$, and the third term scales quadratically.
Notably, a similar structure appears in the standard LLG equation, where both the precession and damping terms are linear in the transverse component of the effective field, $\mathbf{b}_\perp$.
In micromagnetic theory, we study the states close to equilibrium, where $\bm{\nu}$ is nearly aligned with $\bm{\beta}$.
In this regime, $\bm{\beta}_\perp$ is small, and higher-order terms in $\bm{\beta}_\perp$ can be neglected.
Therefore, in the present study, we set $\epsilon = 0$.
A complete treatment of the regularized LLG equation with $\epsilon \neq 0$ will be presented elsewhere.

\vspace{0.5cm}
\noindent
\textbf{External torques}.
\noindent
When external stimuli are applied, we must distinguish whether they are explicitly included in the Hamiltonian \eqref{micro_hamiltonian}, such as a magnetic field gradient, or whether they must be directly incorporated into the LLG equation, for instance, as an electric current.
In the former, the regularized LLG equation \eqref{RLLG} can be used straightforwardly, and in the latter, the corresponding torques have to be added to \eqref{RLLG}.
Here, we consider a practical example of Zhang-Li spin-transfer torque~\cite{Zhang2004}, which is often used to model the magnetization dynamics induced by an electric current.
In case of standard LLG, we have to add to Eq.~\eqref{LLG_micro} the following term:
\begin{equation}
\mathbf{T}=-\dfrac{\xi-\alpha}{1+\alpha^{2}}\mathbf{n}\times\left(\mathbf{u}\cdot\nabla\right)\mathbf{n} + \dfrac{1+\xi\alpha}{1+\alpha^{2}}\left(\mathbf{u}\cdot\nabla\right)\mathbf{n},\label{ZLSL_torque}
\end{equation}
where $\mathbf{u}=\mu_\mathrm{B}\mu_{0}\mathbf{j}/2e\gamma_{0}\mathcal{M}\left(1+\xi^{2}\right)$, $\mathbf{j}$ is the current density vector, and $\xi$ is the non-adiabaticity parameter.
To model this torque in the $\mathbb{S}^{3}$-model, we perform in Eq.~\eqref{ZLSL_torque} a mapping $\mathbf{n}\mapsto\bm{\nu}$ taking into account the constraint $|\bm{\nu}|=1$ and equality $\bm{\nu}\cdot\dot{\bm{\nu}}=0$ that must hold. The resulting expression for the Zhang-Li spin-transfer torque in $\mathbb{S}^{3}$-model can be written as follows:
\begin{equation}
    \!\!\!\bm{\tau}=-\dfrac{\xi-\alpha}{1+\alpha^{2}}\left[\bm{\nu}\times\left(\mathbf{u}\!\cdot\!\nabla\right)\bm{\nu}\times\mathbf{e}_{4}\right]+\dfrac{1+\xi\alpha}{1+\alpha^{2}}\left(\mathbf{u}\!\cdot\!\nabla\right)\bm{\nu},\!\!
    \label{tau_ZL}
\end{equation}
where the expression in square brackets represents the Hodge dual of the wedge product as in Eqs.~\eqref{p_term} and \eqref{f_term}.
A similar approach can be applied to generalize other torques in the regularized LLG equation.

It is worth mentioning that external torques are typically assumed to be small, resulting in ``slow" magnetization dynamics.
Capturing fast dynamical processes requires extending the LLG equation with higher-order terms, such as the nutation term~\cite{Bajpai2019, Mondal2023}, which lies beyond the scope of the present work.
Here, we restrict our analysis to small torques and focus on the near-equilibrium regime, where linear theory is applicable.

\vspace{0.5cm}
\noindent
\textbf{Collective coordinates approach}.
\noindent
The steady-state dynamics of rigid magnetic textures moving at constant velocity $\mathbf{v}$ under external forces or torques is of particular interest in micromagnetic theory~\cite{Malozemoff_79}.
In such cases, the velocity of objects such as domain walls, skyrmions, or vortices can be obtained directly from the Thiele equation~\cite{Thiele_73}, an effective equation of motion derived using the collective coordinate approach.
This method avoids solving the full LLG equation and eliminates the need for time-consuming micromagnetic simulations.
Moreover, the solution of the Thiele equation can be considered an exact result toward which numerical simulations should converge in the steady-state limit.
This makes it a valuable benchmark for verifying the correctness of numerical implementations and demonstrating the internal consistency of the micromagnetic theory.
Below we derive a generalized version of the Thiele equation for the $\mathbb{S}^{3}$-model.

First, we parametrize $\bm{\nu}$ using spherical coordinates on the three-sphere $\mathbb{S}^{3}$, denoted by angles $\Theta$, $\Phi$, and $\Psi$:
\begin{equation}
\bm{\nu} = 
\begin{pmatrix}
\sin\Theta \cos\Phi \cos\Psi \\
\sin\Theta \sin\Phi \cos\Psi \\
\cos\Theta \cos\Psi \\
\sin\Psi
\end{pmatrix}^{\!\!T}
\end{equation}
Using this parametrization, we reformulate the regularized LLG equation~\eqref{RLLG} in terms of the spherical angles $\Theta$, $\Phi$, and $\Psi$ (see Supplementary Material~III).
In the case of rigid motion of a magnetic texture, the following relation holds: $\mathcal{R}(\mathbf{r}, t) = \mathcal{R}(\mathbf{r} - \mathbf{v}t)$ for $\mathcal{R} \in {\Theta, \Phi, \Psi}$.
This implies that the time derivatives of the spherical angles are given by:
\begin{equation}
\dot{\Theta} = -\mathbf{v} \cdot \nabla\Theta, \quad
\dot{\Phi} = -\mathbf{v} \cdot \nabla\Phi, \quad
\dot{\Psi} = -\mathbf{v} \cdot \nabla\Psi.
\label{TFP_v}
\end{equation}
The left-hand sides of these equations can be directly obtained from the regularized LLG equation derived in the previous step.
Following the standard procedure~\cite{Malozemoff_79}, we next assume that during steady motion the total energy of the system remains constant.
As a result, the dissipation function must vanish:
\begin{equation}
\displaystyle\int\left(\dfrac{\delta\mathcal{E}}{\delta\Theta}\nabla\Theta+\dfrac{\delta\mathcal{E}}{\delta\Phi}\nabla\Phi+\dfrac{\delta\mathcal{E}}{\delta\Psi}\nabla\Psi\right)\mathrm{d}V=0.\label{Force}
\end{equation}
The combined solution of Eqs.~\eqref{TFP_v} and \eqref{Force} yields the Thiele equation (see Supplementary Material~IV):
\begin{equation}
\alpha\, \mathbf{g} \times \mathbf{v} + \hat{\gamma} \mathbf{v} = \mathbf{f}, \label{Thiele}
\end{equation}
where the force due to the electric current is given by $\mathbf{f}=-\alpha\mathbf{g}\times\mathbf{u}-\left[\left(1+\xi\alpha\right)\hat{\gamma} +\left(\xi-\alpha\right)\hat{\gamma}^{\prime}\right]\mathbf{u}/\left(1+\alpha^2\right)$.
Here, $\mathbf{g}$ is the gyrovector, and $\hat{\gamma}$ and $\hat{\gamma}^{\prime}$ are the dissipation tensors, whose components are defined as follows:
\begin{eqnarray}
    && g_{i}=\epsilon_{ijk}\intop\dfrac{\mathbf{n}\cdot\partial_{j}\mathbf{n}\times\partial_{k}\mathbf{n}}{n^{2}+\alpha^{2}}\mathrm{d}V,\,\,\{i,j,k\}\in\{x,y,z\},\nonumber\\
&&\hat{\gamma}_{jk}=\intop\dfrac{\alpha^{2} \partial_{j}\bm{\nu}\cdot\partial_{k}\bm{\nu}+\partial_{j}\nu_{4}\partial_{k}\nu_{4}}{n^{2}+\alpha^2}\mathrm{d}V,\\
&&\hat{\gamma}_{jk}^{\prime}=\alpha\intop\dfrac{n^{2}\partial_{j}\bm{\nu}\cdot\partial_{k}\bm{\nu}-\partial_{j}\nu_{4}\partial_{k}\nu_{4}}{n^{2}+\alpha^2}\mathrm{d}V\nonumber.\label{gyro_diada}
\end{eqnarray}
where $\epsilon_{ijk}$ is the Levi-Civita symbol.
The derived Eq.~\eqref{Thiele} is applicable for electric currents $\mathbf{u}$ applied in arbitrary directions when the background magnetization is uniform, i.e., in a saturated state.
However, when the vacuum corresponds to a non-uniform magnetic configuration, such as a spin spiral with wave vector $\mathbf{k}$, natural constraints on the current direction arise.
In particular, to avoid excitation of the background magnetization, the current must lie in the plane orthogonal to the wave vector, i.e., $\mathbf{u} \perp \mathbf{k}$.
These and related constraints on the motion of three-dimensional magnetic textures in non-uniform backgrounds were analyzed in detail in Ref.~\cite{Kuchkin2022}.

In the following examples, we set $\mathbf{k} \parallel \mathbf{e}_z$ and $\mathbf{u} \parallel \mathbf{e}_x$, so the current does not excite the background magnetization.
Consequently, for the skyrmion string and the chiral bobber, the soliton velocity has no out-of-plane component: $\mathbf{v} = (v_x, v_y, 0)$.
For fully 3D magnetic solitons -- such as hopfions~\cite{Liu2020}, dipolar strings~\cite{Kuchkin2025_2}, and hybrid skyrmion tubes~\cite{Kuchkin2022} -- this restriction no longer applies, allowing solitons to move in any direction. 
A general solution to the Thiele equation for this case is derived in Supplementary Material~V.

Finally, it is worth noting that setting $\nu_{4}=0$ in Eq. \eqref{Thiele} reduces the equation to the standard Thiele equation:
\begin{equation}
\mathbf{G}\times\mathbf{v}+\alpha\Gamma\mathbf{v}=\mathbf{F},\label{classical_Thiele}
\end{equation}
where the gyro-vector $\mathbf{G}$ has components $G_{i}=\int\left[\epsilon_{ijk}\mathbf{n}\cdot\partial_{j}\mathbf{n}\times\partial_{k}\mathbf{n}\right]\mathrm{d}V$, dissipation tensor $\Gamma$ has components $\Gamma_{jk}=\int\left[\partial_{j}\mathbf{n}\cdot\partial_{k}\mathbf{n}\right]\mathrm{d}V$, and $\mathbf{F}$ is an external force, which for the case of Zhang-Li torque is given by $\mathbf{F}=-\mathbf{G}\times\mathbf{u}-\xi\Gamma\mathbf{u}$.

\vspace{0.5cm}
\noindent
\textbf{Current-induced soliton dynamics}.
In this section, we present the results of numerical simulations on the dynamics of BP-hosting spin textures using the regularized LLG equation~\eqref{RLLG} with the Zhang-Li torque~\eqref{tau_ZL}.
These results are compared with simulations based on the standard LLG simulations in Mumax3~\cite{Mumax}.
Our implementation of the regularized LLG equation is available in a public repository~\cite{rllg_git}, which represent a modified fork of the Mumax3 code.
In addition to the regularized LLG equation, this fork includes several other advanced features.
For example, it provides an implementation of the regularized geodesic nudged elastic band method for the $\mathbb{S}^3$ model, which we previously used in Ref.~\cite{Kuchkin2025_2}.

To illustrate the advantages of the regularized over the standard LLG equation, we consider a skyrmion tube [Fig.~\ref{Fig1}\textbf{c}], chiral bobber~\cite{Rybakov2015} [Fig.~\ref{Fig1}\textbf{d}], and dipole string~\cite{Kuchkin2025_2} [Fig.~\ref{Fig1}\textbf{e}].
These solitons are representative examples of magnetic textures containing zero, one and two BPs, respectively.
These solitons are stabilized in chiral magnets, where the natural background (vacuum) state is a helical or conical phase.
We consider a system of size $2L_\mathrm{e} \times 2L_\mathrm{e} \times 2L_\mathrm{e}$, where $L_\mathrm{e}$ is the equilibrium period of chiral modulations.
The material parameters used to stabilize the skyrmion string and the chiral bobber are identical, whereas the dipolar string requires slightly different conditions (see Methods).
For the skyrmion string and chiral bobber, we apply periodic boundary conditions in the $xy$-plane and open boundary conditions along the $z$-axis.
In the case of the dipolar string, a fully 3D magnetic soliton, we model a bulk crystal and impose periodic boundary conditions in all three directions.
In our simulations, we use discrete meshes of varying densities to demonstrate how the cuboid size influences Bloch point dynamics in both models (see legend in Fig.~\ref{Fig1}\textbf{d}).

In the case of the skyrmion string [Fig.~\ref{Fig1}\textbf{d}, \textbf{g}], simulations using both the standard and regularized LLG equations yield identical results. 
In both models, the skyrmion moves with the same deflection angle $\arctan (v_y/v_x)$.
Importantly, the outcomes of these simulations are stable with respect to varying mesh density and show excellent agreement with analytical predictions from the Thiele equation (see Supplementary Table 1).
In particular, as the current density $\mathbf{j}$ decreases, both the longitudinal ($v_x$) and transverse ($v_y$) components of the velocity decrease linearly and vanish in the limit $\mathbf{j} \rightarrow 0$.

In contrast to the skyrmion string case, the results for the chiral bobber and dipole string reveal substantial discrepancies between the standard and regularized micromagnetic models.
In the standard $\mathbb{S}^2$-based model, the velocity exhibits a nonlinear dependence on current density, with a nonzero critical current below which motion does not occur [Fig.~\ref{Fig1}\textbf{e} and \textbf{f}].
This behavior cannot be justified within the framework of continuum theory, where the magnetic medium is treated as continuous and free from intrinsic thresholds for magnetization dynamics.
It also contradicts the Thiele equation, which predicts a linear dependence of velocity on current density.
These artifacts indicate a fundamental failure of the standard micromagnetic model to correctly describe systems containing magnetic singularities.
Moreover, simulations based on the $\mathbb{S}^2$-model do not exhibit convergence with increasing mesh density, which makes them unreliable for studying Bloch point dynamics.

The most striking inconsistency observed in simulations of the chiral bobber and dipole string is the inverted skyrmion Hall angle upon varying current density and mesh discretization [see sign reversal of $v_{y}$ in Fig.~\ref{Fig1}\textbf{e} and \textbf{f}].
Such behavior of the skyrmion Hall angle in micromagnetic simulations contradicts the Thiele equation~\eqref{classical_Thiele} and thus represents an artifact. 
In continuum models, the sign of the skyrmion Hall angle is uniquely determined by the topological charge of the magnetic texture and remains invariant.

On the other hand, the regularized $\mathbb{S}^3$-based model consistently produces physically meaningful results that do not show noticeable dependence on the mesh density [Fig.~\ref{Fig1}\textbf{h} and \textbf{i}].
Similar to the case of skyrmion string [Fig.~\ref{Fig1}\textbf{g}], the velocities linearly converge to zero with current density, as predicted by the Thiele equation ~\eqref{Thiele} (see Supplementary Table 1).

\begin{figure*}[tb!]
\centering
\includegraphics[width=17.4cm]{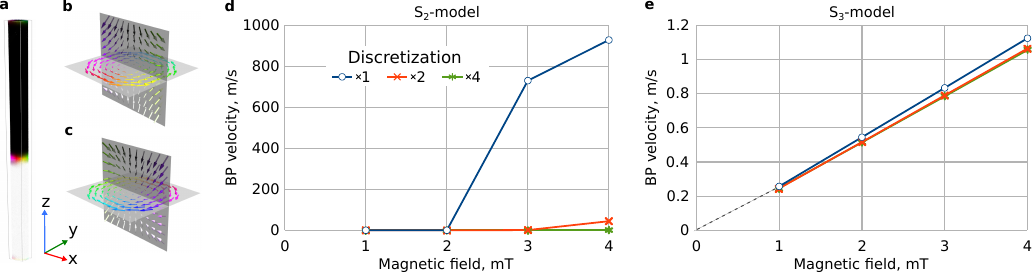}
\caption{~\small \textbf{BP dynamics in a nanowire}. 
\textbf{a} A magnetic nanowire containing two domains magnetized along $+\mathbf{e}_z$ (white) and $-\mathbf{e}_z$ (black), which are separated by a domain wall containing a Bloch point (BP).
In an external magnetic field $\mathbf{B}_\mathrm{ext} \parallel \mathbf{e}_z$, the BP moves along the wire axis.
Simulations were performed for two domain-wall configurations, shown in \textbf{b} and \textbf{c}.
Both configurations exhibit identical BP dynamics.
\textbf{d} and \textbf{e} show the BP velocity as a function of the external magnetic field strength, computed using the standard and regularized LLG equations, respectively.
Simulations were carried out at different discretization levels:
$\times$1 corresponds to a $16 \times 16 \times 512$ mesh, while $\times$2 and $\times$4 indicate uniform scaling of each spatial dimension by factors of 2 and 4, respectively.
Domain size and material parameters are provided in the Methods section.
} 
\label{Fig2}
\end{figure*}

\vspace{0.5cm}
\noindent
\noindent
\textbf{BP motion in a nanowire}. In magnetic nanowires, a BP can appear at the center of a domain wall separating two oppositely magnetized domains aligned along the wire axis [Fig.~\ref{Fig2}\textbf{a}].
Such domain walls arise from the competition between the Heisenberg exchange interaction and the demagnetizing field.
They have been extensively studied both experimentally~\cite{Fert1999, Biziere2013, DaCol2014, Schbitz2019, wartelle_19, Donnelly2024}
and theoretically~\cite{Hertel2004, Thiaville, Hertel2016,  Charilaou2018,  Ma2020, Charilaou2022, Moreno2022, FernandezRoldan2022, DualBP2022, FernandezRoldan2020, Tejo2024, Bittencourt2024, Saugar2025}.
In the following, we consider a soft magnetic material modeled by taking into account the Heisenberg exchange, demagnetizing field, and Zeeman energy terms.
Under these conditions, the head-to-head domain wall structure depicted in Fig.~\ref{Fig2}\textbf{a} admits two energetically equivalent BP configurations, as shown in Figs.~\ref{Fig2}\textbf{b} and \textbf{c}.
A key feature of these systems is that a weak external magnetic field can readily induce BP dynamics.
The domain whose magnetization is aligned with the field expands, causing the domain wall and the enclosed BP to move along the wire.
Here, we focus on the low-field, linear-response regime, where the dynamics can be well approximated by the rigid motion of the spin texture.
Nonlinear effects, such as those associated with ultrafast domain wall motion~\cite{Tejo2024}, are beyond the scope of this work.

Figures~\ref{Fig2}\textbf{d} and \textbf{e} show simulation results for BP dynamics under external fields ranging from 1~mT to 4~mT, using the standard and regularized micromagnetic models, respectively.
The results from the standard LLG simulations reveal unphysical behavior.
In particular, the response of the domain wall to an applied magnetic field exhibits oscillatory behavior.
As shown in Supplementary Figure 1, the temporal evolution of the magnetization components ($n_x$, $n_y$, $n_z$) exhibits oscillations with frequencies that cannot be explained within a continuum approximation.
More importantly, Fig.~\ref{Fig2}\textbf{d} shows that increasing mesh density leads to a suppression of BP motion.
Below a certain discretization threshold, the BP becomes fully pinned and stops moving.
This implies that an unphysical numerical parameter -- the mesh density -- determines the behavior of the system.
Equivalently, we can say that in the standard micromagnetic model, there exists a mesh density-dependent critical field below which BP does not move.
Thereby, we conclude that such a pinning field is a numerical artifact, arising from the divergence of the effective field at the BP core in the standard micromagnetic model.

Although the results in Fig.~\ref{Fig2}\textbf{d} correspond to a specific parameter set, the observed BP pinning effect—whether due to increasing mesh resolution or decreasing external field—is a general artifact inherent to the standard micromagnetic model.

In contrast, the regularized $\mathbb{S}^3$-model is free from this issue [Fig.~\ref{Fig2}\textbf{e}].
Like the chiral bobber and dipolar string, the BP in this model exhibits smooth motion, with velocity continuously tending to zero as $B_\mathrm{ext} \rightarrow 0$.
Moreover, the BP dynamics in the regularized model do not induce any artificial magnetization oscillations (see Supplementary Figure 1).

\vspace{0.5cm}
\noindent
\textbf{Discussions} 

\noindent
\textbf{On the $\kappa$-parameter}.
Micromagnetic theory operates on several characteristic length scales that depend on the material type and the dominant interactions.
In uniaxial ferromagnets, this length is typically defined by the domain wall width~\cite{hubert2008magnetic},
$L_\mathrm{c} = \sqrt{2\mathcal{A} / \left(2\mathcal{K} + \mu_0 \mathcal{M}^2\right)}$.
In isotropic chiral magnets, the characteristic length corresponds to the equilibrium period of the spin spiral~\cite{Bogdanov1989},
$L_\mathrm{D} = 4\pi\mathcal{A} / \mathcal{D}$,
where $\mathcal{D}$ denotes the DMI constant.
In the case of textures such as chiral kinks~\cite{Kuchkin2020}, a shorter characteristic length appears,
$L_\mathrm{ck} \sim 0.1 L_\mathrm{D}$.
In exchange-frustrated systems, the characteristic scale is determined by the ratio of competing exchange terms~\cite{Rybakov2022}.

In the regularized micromagnetic model, the size of the magnetic texture surrounding a point singularity (the “size” of the Bloch point) defines an additional characteristic length,
$L_\kappa = \sqrt{\mathcal{A} / \kappa}$.
To ensure the accuracy of numerical simulations, the discretization grid must be much smaller than the shortest relevant characteristic length.
In the examples presented in the main text, $\kappa$ was chosen such that $L_\kappa$ is comparable to either $L_\mathrm{c}$ or $L_\mathrm{D}$.
In general, however, $\kappa$ is determined by the properties of the specific material (lattice symmetry, sort of atoms, electron density of states, etc.) and does not necessarily correlate with the other characteristic lengths.

\noindent
\textbf{Anisotropic systems}.
In the most general case, the exchange stiffness $\mathcal{A}$ is a second-rank symmetric tensor, which reduces to a scalar only in isotropic systems.
In anisotropic systems, such as certain hexagonal crystals~\cite{grundy1972BubblesCobalt}, multilayered structures~\cite{Beg2019,Cocoons_22,Lang2023}, or van der Waals magnets \cite{Han2019}, the Bloch point may lose its spherical symmetry.
As a result, the characteristic size of the magnetic point singularity becomes direction-dependent.

To determine the appropriate value of the regularization parameter $\kappa$ for a given material, one must rely on experimental data or microscopic calculations based on more fundamental models, such as the quantum Heisenberg model or density functional theory~\cite{Szilva2023}.
While the design proposal of an experiment reaches beyond the scope of this study, we want to note that neutron scattering cross sections are sensitive to the norm of the magnetization~\cite{Kuchkin2025}, which could prove useful to estimate $\kappa$.

\noindent
\textbf{Classical spin lattice models}.
Magnetism is inherently of quantum mechanical origin, and quantum models offer the most fundamental framework for describing magnetic materials. 
However, the complexity of these models limits their applicability to only the simplest systems.
In practice, we often have to rely on simplified models.

One such model is the quantum Heisenberg model, which treats spins as quantum objects but uses a simplified Hamiltonian.
While this approach provides valuable insights~\cite{Haller2022, Rzsa2025}, it is computationally limited to systems of about a few thousand spins.
In contrast, real magnetic systems typically contain billions of spins.

The next level of simplification is classical spin lattice models.
They can reproduce many experimental results and extend applicability to systems on the order of tens of nanometers.
The applicability of classical atomistic spin models, where the magnetic moment has a fixed length, is justified only in certain limiting cases~\cite{Millard1971}, and is generally regarded as a useful but simplified approximation.

To reach micrometer scales, one usually employs the micromagnetic theory, which describes the classical spin lattice model in the continuum limit.
Although its predictive power, micromagnetics inherits the limitations of classical spin lattice models and imposes an additional constraint of magnetic texture continuity. 
Most importantly, both approaches neglect the intrinsic quantum nature of spin. 
Various multiscale approaches~\cite{Hertel2015}, which aim to combine standard micromagnetic and atomistic spin lattice models, offer a promising framework but do not overcome the fundamental limitations discussed above.
Due to their inherent complexity, such methods remain less thoroughly examined in terms of agreement with experimental observations and may introduce additional artifacts at the borders between the regions described by different Hamiltonians.

The regularized micromagnetic model introduced here is not derived from the classical fixed-length spin lattice model and therefore avoids its inherent limitations.
Instead, it accounts for the possibility of magnetization length reduction inherently present in more general quantum spin Hamiltonians.
Accordingly, it is more appropriate to compare our model with such quantum atomistic models rather than classical spin models.
Thereby, for the magnetic textures with singularities, our model is not expected to converge to classical atomistic spin dynamics.
However, for smooth magnetic textures such as vortices, skyrmions, and hopfions, the regularized model agrees well with both classical micromagnetics and classical atomistic simulations.

\noindent
\textbf{Temperature in micromagnetism}. According to the definition given by Landau, micromagnetism is an \textit{athermal} theory~\cite{hubert2008magnetic}.
It means that temperature enters the model Hamiltonian only implicitly --- through the temperature dependence of material parameters such as saturation magnetization $\mathcal{M}$, anisotropy tensor $\hat{K}$, exchange stiffness $\mathcal{A}$, and DMI constant $\mathcal{D}$.
This reflects a key point: as long as the system exhibits a magnetic order, regardless of the particular temperature, the Hamiltonian~\eqref{micro_hamiltonian} remains valid for describing its physical behavior.
On the other hand, as the temperature approaches the Curie point $T_\mathrm{c}$, the model may fail to capture effects driven by entropy and thermal fluctuations.
However, as long as $T \ll T_\mathrm{c}$, the entropy contribution to the free energy is assumed to be minimal. 
For most practical problems addressed by micromagnetics, it is sufficient to assume that the material parameters in~\eqref{micro_hamiltonian} are functions of temperature.

At elevated temperatures near the Curie point ($T \simeq T_\mathrm{c}$), the magnitude of the saturation magnetization can be modeled using a Landau expansion~\cite{Galkina1993}, $e_L = a n^2 + b n^4$, which captures the characteristics of a second-order phase transition~\cite{Landau1937}.
In this regime, two well-established dynamical equations have been proposed: the Landau–Lifshitz–Bloch (LLB) equation~\cite{Garanin1997}, derived from the Fokker–Planck formalism, and the phenomenological Landau–Lifshitz–Baryakhtar (LLBar) equation~\cite{Baryakhtar1984}.
Both frameworks serve as generalizations of the LLG equation, incorporating additional terms that allow for temporal variations in the magnitude of the magnetization.
This leads to a natural question: can these models be used to describe Bloch point (BP) dynamics by treating the temperature-dependent parameters as free constants, even without considering temperature effects?
Below, we outline several reasons why this approach is not suitable.

First, the Landau energy term does not inherently enforce the constraint $|\mathbf{n}| \leq 1$. 
As a result, the model permits unphysical increases in the magnetization magnitude beyond saturation. 
Regardless of how rare such deviations may be, the mere possibility of violating this fundamental constraint limits the applicability of the model to a narrow parameter range.

Second, there is no established method for incorporating external torques into the LLB or LLBar equations while preserving the constraint $|\mathbf{n}| \leq 1$, which is crucial for physically consistent modeling. 
In contrast, the approach developed in this work naturally accommodates such torques. 
We explicitly demonstrate this by including the spin-transfer torque induced by spin-polarized electric currents.

Third, the experimentally observed dynamics of Bloch point–hosting spin textures (e.g., hard magnetic bubbles) are well described by the Thiele equation. 
This equation can be derived from the regularized LLG framework proposed here, but not from the LLB or LLBar equations.

In conclusion, there is no clear advantage in employing the Landau energy term and LLB or LLBar dynamics for modeling Bloch point behavior. 
The $\mathbb{S}^3$-based formulation offers a more consistent and physically grounded framework.

\noindent
\textbf{Methods}

\noindent
\textbf{Micromagnetic simulations of soliton dynamics}.
\noindent
To stabilize the skyrmion tube, chiral bobber, and dipole string (Fig.~\ref{Fig1}(c)-(e)), we consider the following Hamiltonian for chiral magnets:
\begin{equation}
\mathcal{E}_{1}=\intop\left[e_\mathrm{exi}+e_\mathrm{dmi}+e_\mathrm{ani}+e_\mathrm{z}\right]\mathrm{d}V,
\end{equation}
with the following parameters: $\mathcal{A}=4$ pJ/m, $\mathcal{D}=0.718$ mJ/m$^2$, $\mathcal{M}=384$ kA/m$^{3}$. 
These parameters correspond to a spin-spiral period of $L_\mathrm{D}=70$ nm. 
We further assume the $\kappa=10^{-4}\mathcal{D}^{2}/2\mathcal{A}$ which corresponds to a characteristic BP size of $L_{\kappa}\approx 7.9$ nm.
To stabilize skyrmion and chiral bobber, we apply an external field of $0.7\mathcal{D}^{2}/2\mathcal{A}\mathcal{M}$, which corresponds to the equilibrium period of spin-spiral $L_\mathrm{e}=L_\mathrm{D}$.
To stabilize dipole string we apply an external field of $0.55\mathcal{D}^{2}/2\mathcal{A}\mathcal{M}$ and additionally add easy-plane anisotropy of strength $K_{\mathrm{u},xx}=-0.25\mathcal{D}^{2}/2\mathcal{A}$ (hard axis along $x$), that results in  $L_\mathrm{e}\approx 1.016 L_\mathrm{D}$.
The soliton motion was excited by electric current of density $\mathbf{j}=j\mathbf{e}_{x}$ of various strengths, $\alpha=0.05$ and $\xi=0.25$.
Numerical simulations were performed using the Mumax3 software for which we implemented the regularized LLG equation \eqref{RLLG} with Zhang-Li torque \eqref{tau_ZL}~\cite{rllg_git}.

\noindent
\textbf{Semi-analytical Thiele approach}. To compare the results of micromagnetic simulations with the predictions of the Thiele equation \eqref{Thiele}, we extracted information about skyrmion velocities from the simulation data.
The position of soliton's center, $\mathbf{r}_\mathrm{c}$, taking into account the PBC was calculated as follows~\cite{Kuchkin2021, Kuchkin2025_2}:
\begin{align}
    r_{i,\mathrm{c}}\!=\!\dfrac{L_{i}}{2\pi}\tan^{-1}\!\dfrac{\int\!\mathcal{N}_{jk}\sin\left(2\pi r_{i}/L_{i}\right)\mathrm{d}r_{i}}{\int\!\mathcal{N}_{jk}\cos\left(2\pi r_{i}/L_{i}\right)\mathrm{d}r_{i}}\pm l_{i}L_{i},\label{center_mass}
\end{align}
where the non-repeating indices $\{i,j,k\}\in\{x,y,z\}$, and $\mathcal{N}_{jk}=\int(1-n_z)\mathrm{d}r_{j}\mathrm{d}r_{k}$, $L_{x}$, $L_{y}$ and $L_{z}$ is the size of simulation domain.
The integers $l_{i}$ represent the number of times the soliton has crossed the domain boundary in the $x$, $y$, and $z$ directions, respectively.
Both solutions of the Thiele equation and numerical simulations shown in Fig.~\ref{Fig1} suggest a linear dependency between solitons' velocity components and the current density, $v_{i}=-c_i{u}$, where dimensionless coefficient $c_{i}$ is given by the solution of the Thiele equation.
In Supplementary Table 1, we provide numerical values of this coefficient.

\noindent
\textbf{Micromagnetic simulations of BP in a nanowire}. We considered an Fe nanowire of length $1.2 \,\mu$m and of diameter $35$ nm which can be described by the Hamiltonian:
\begin{equation}
    \mathcal{E}_{2}=\intop\left[e_\mathrm{exi}+e_\mathrm{ddi}+e_\mathrm{z}\right]\mathrm{d}V,
\end{equation}
with $\mathcal{A}=21$ pJ/m and $\mathcal{M}=1.7$ MA/m. 
For the $\mathbb{S}^3$-model, we set $\kappa=0.5$ MJ/m$^3$ that corresponds to a characteristic BP size of $L_{\kappa}=\sqrt{\mathcal{A}/\kappa} = 42$ nm.
The initial configuration shown in Fig.~\ref{Fig1}\textbf{a} is stabilized at zero external field.
To initiate domain wall motion, a magnetic field is applied along the wire, $B_\mathrm{ext} \parallel \mathbf{e}_z$.
In both the standard and regularized LLG simulations, we use the same damping parameter, $\alpha = 0.01$.
The position of the Bloch point, $z_\mathrm{p}$, is estimated as:
\begin{equation}
    z_\mathrm{p}=\dfrac{\int r_{z}\left[n_{x}^{2}+n_{y}^{2}\right]\mathrm{d}V}{\int\left[n_{x}^{2}+n_{y}^{2}\right]\mathrm{d}V},\label{pos_z}
\end{equation}
that always provides us with a finite value as long as the domain wall is present in the wire.

\noindent
\textbf{Acknowledgments}
Nikolai S.\ Kiselev and Vladyslav M.\ Kuchkin acknowledge F.\ N.\ Rybakov for fruitful discussions and critical comments.
Vladyslav M.\ Kuchkin, Andreas Haller, Thomas Schmidt, and Andreas Michels acknowledge financial support from the National Research Fund of Luxembourg (PRIDE MASSENA Grant, AFR Grant No.~15639149, and DeQuSky Grant No.~C22/MS/17415246).
Vladyslav M.\ Kuchkin acknowledges financial support from the European Union’s Horizon 2020 research and innovation programme under the Marie Sk{\l}odowska-Curie grant agreement No.~101203692 (QUANTHOPF).
Nikolai S.\ Kiselev acknowledges support from the European Research Council under the European Union's Horizon 2020 Research and Innovation Programme (Grant No.~856538---project ``3D MAGiC'').

\clearpage
%%%%%%%%%%%%%%%%%%%%%%%%%%%%%%
\bibliographystyle{apsrev4-2}
\bibliography{main}

\newpage

\onecolumngrid
% \appendix
\newpage
\begin{center}
   \textbf{Supplemental Material for ``Regularized Micromagnetic Theory for Bloch Points''}\newline 
\end{center}

\section{Regularization of the effective field for spherically symmetric Bloch point}\label{A:regularizartion}
For the spherically symmetric case, $\mathbf{n}=\mathbf{r}/r$, let us check whether it is possible to regularize the effective field $\mathbf{b}_\mathrm{exi}=-2\mathcal{A}\nabla^{2}\mathbf{n}$ when $\mathbf{n}=n_{0}\mathbf{r}/r$ for some $n_{0}=n_{0}(r)$.
We can find in this case,
\begin{eqnarray}
    \nabla^{2}\mathbf{n}=\left[\nabla^{2} n_{0}-2\dfrac{n_{0}}{r^{2}}\right]\dfrac{\mathbf{r}}{r}.\label{triangl_n}
\end{eqnarray}
Assuming that $n_{0}(r)$ is continuous,  it can be expanded as a Taylor series:
\begin{equation}
    n_{0}\left(r\right)=\sum_{n=0}^{\infty}c_{n}r^{n}.\label{n0_series}
\end{equation}
Then plugging \eqref{n0_series} into \eqref{triangl_n}, we get
\begin{eqnarray}
    &&\nabla^{2}\mathbf{n}=\left[2c_{2}+6c_{3}r+12c_{4}r^{2}+\mathcal{O}\left(r^{3}\right)-2\dfrac{c_{0}+c_{1}r+c_{2}r^{2}+c_{3}r^{3}+c_{4}r^{4}+\mathcal{O}\left(r^{5}\right)}{r^{2}}\right]\dfrac{\mathbf{r}}{r}\nonumber\\
    &&=\left[-\dfrac{2c_{0}}{r^{2}}-\dfrac{2c_{1}}{r}+4c_{3}r+10c_{4}r^{2}+\mathcal{O}\left(r^{3}\right)\right]\dfrac{\mathbf{r}}{r}.\label{n2}
\end{eqnarray}
Notably, this expression does not contain a constant term (i.e. term next to $r^{0}$).
The only way to ensure RHS of \eqref{n2} remains finite at $r\rightarrow0$ is to require $c_{0}=c_{1} = 0$. 
This implies, $n_{0}(0)=0$.

\section{A proof that $\mathbf{h}$ can not be linear in components of the effective field}\label{A:h_proof}

By construction, precession $\mathbf{p}$ and dissipation $\mathbf{d}$ vectors are linear in terms of the components of the effective field $\bm{\beta}$, while the expression for $\mathbf{h}=\bm{\nu}\times\mathbf{p}\times\mathbf{d}$ is in fact quadratic in $\bm{\beta}$.
Let us try to construct $\mathbf{h}$ as a linear form instead assuming that $|\bm{\nu}\cdot\bm{\beta}|>0$ and $|\bm{\beta}|>0$ corresponding to out-of-equilibrium regime.

First of all, due to $|\bm{\beta}|>0$, we can always choose an orthonormal basis $\left\{\bm{l}_{1},\bm{l}_{2},\bm{l}_{3},\bm{l}_{4}\right\}$ in $\mathbb{R}^{4}$ associated with $\bm{\beta}$:
\begin{eqnarray}
    \bm{l}_{1}=\bm{\beta}/|\bm{\beta}|.
\end{eqnarray}
Then spanning $\mathbf{h}$ on this basis:
\begin{eqnarray}
\mathbf{h}=b_{1}\bm{l}_{1}+b_{2}\bm{l}_{2}+b_{3}\bm{l}_{3}+b_{4}\bm{l}_{4},
\end{eqnarray}
and remembering that $\mathbf{h}$ lies in the tangent space of $\bm{\nu}$, i.e. $\mathbf{h}\cdot\bm{\nu}=0$, we can check for which coefficients $b_{1}, b_{2}, b_{3}, b_{4}$, it holds $\mathbf{h}\cdot\mathbf{d}=0$:
\begin{eqnarray}
    \mathbf{h}\cdot\mathbf{d}=\mathbf{h}\cdot\left[\bm{\nu} \left(\bm{\nu} \cdot \bm{\beta} \right) - \bm{\beta}\right]=-\left(b_{1}\bm{l}_{1}+b_{2}\bm{l}_{2}+b_{3}\bm{l}_{3}+b_{4}\bm{l}_{4}\right)\cdot\bm{\beta}=-b_{1}\bm{l}_{1}\cdot\bm{\beta}=-b_{1}|\bm{\beta}|=0.
\end{eqnarray}
For $|\bm{\beta}|>0$, the last equality holds only when $b_{1}=0$. Thus, projection of $\mathbf{h}$ onto the vector $\bm{\beta}$ is zero, or equivalently $\mathbf{h}\perp\bm{\beta}$. 

For given two vectors $\bm{\nu}$ and $\bm{\beta}$ with $|\bm{\nu}\cdot\bm{\beta}|>0$, in $\mathbb{R}^{4}$ it is possible to construct only two linearly independent vectors orthogonal to both $\bm{\nu}$ and $\bm{\beta}$.  
For instance, the first vector can be chosen as $\mathbf{p}=\bm{\nu}\times\bm{\beta}\times\mathbf{e}_{4}$, then the second vector ($\mathbf{h}$) can be defined as:
\begin{equation}
    \mathbf{h}=c_{1}\bm{\nu}\times\bm{\beta}\times\mathbf{e}_{1}+c_{2}\bm{\nu}\times\bm{\beta}\times\mathbf{e}_{2}+c_{3}\bm{\nu}\times\bm{\beta}\times\mathbf{e}_{3}.\label{h_expr}
\end{equation}
Note, that term $c_{4}\bm{\nu}\times\bm{\beta}\times\mathbf{e}_{4}$ is present in $\mathbf{p}$ and its inclusion in $\mathbf{h}$ just leads to redefinition of the coefficient $a_{1}$ in the regularized LLG equation.
Note that \eqref{h_expr} is the most general form of $\mathbf{h}$, where $c_{1}$, $c_{2}$ and $c_{3}$ are nonzero. 
However, it is sufficient to consider only one of these coefficients as nonzero.
Moreover, one can check that four vectors $\left\{\bm{\nu}\times\bm{\beta}\times\mathbf{e}_{1}, \bm{\nu}\times\bm{\beta}\times\mathbf{e}_{2}, \bm{\nu}\times\bm{\beta}\times\mathbf{e}_{3}, \bm{\nu}\times\bm{\beta}\times\mathbf{e}_{4}\right\}$ are coplanar and lie in a two-dimensional subspace of four-dimensional space, $\mathbb{R}^{2} \subset \mathbb{R}^{4}$, which is orthogonal to the plane spanned by $\bm{\nu}$ and $\bm{\beta}$.
Now let us check for which coefficients $c_{1}, c_{2}, c_{3}$ it holds $\mathbf{h}\cdot\mathbf{p}=0$,
\begin{eqnarray}
    &&\!\!\!\mathbf{h}\cdot\mathbf{p}=\left(c_{1}\bm{\nu}\times\bm{\beta}\times\mathbf{e}_{1}+c_{2}\bm{\nu}\times\bm{\beta}\times\mathbf{e}_{2}+c_{3}\bm{\nu}\times\bm{\beta}\times\mathbf{e}_{3}\right)\cdot\mathbf{p}=\left[\left(\nu_{3}\beta_{4}-\nu_{4}\beta_{3}\right)c_{2}-\left(\nu_{2}\beta_{4}-\nu_{4}\beta_{2}\right)c_{3}\right]\left(\nu_{2}\beta_{3}-\nu_{3}\beta_{2}\right)\nonumber\\
    &&+\left[\left(\nu_{1}\beta_{4}-\nu_{4}\beta_{1}\right)c_{3}-\left(\nu_{3}\beta_{4}-\nu_{4}\beta_{3}\right)c_{1}\right]\left(\nu_{3}\beta_{1}-\nu_{1}\beta_{3}\right)+\left[\left(\nu_{2}\beta_{4}-\nu_{4}\beta_{2}\right)c_{1}-\left(\nu_{1}\beta_{4}-\nu_{4}\beta_{1}\right)c_{2}\right]\left(\nu_{1}\beta_{2}-\nu_{2}\beta_{1}\right).
\end{eqnarray}
This expression is nothing but a quadratic form:
\begin{equation}
    \mathbf{h}\cdot\mathbf{p}=\bm{\beta}^{T}A\bm{\beta},
\end{equation}
where $A$ is a symmetric matrix ($A_{ij}=A_{ji}$) given by:
\begin{equation}
    A=\dfrac{1}{2}\left(\begin{array}{cccc}
-2\nu_{4}\left(c_{3}\nu_{3}+c_{2}\nu_{2}\right) & \nu_{4}\left(c_{1}\nu_{2}+c_{2}\nu_{1}\right) & \nu_{4}\left(c_{1}\nu_{3}+c_{3}\nu_{1}\right) & \nu_{1}\left(c_{2}\nu_{2}+c_{3}\nu_{3}\right)-c_{1}\left(\nu_{2}^{2}+\nu_{3}^{2}\right)\\
* & -2\nu_{4}\left(c_{3}\nu_{3}+c_{1}\nu_{1}\right) & \nu_{4}\left(c_{2}\nu_{3}+c_{3}\nu_{2}\right) & \nu_{2}\left(c_{1}\nu_{1}+c_{3}\nu_{3}\right)-c_{2}\left(\nu_{1}^{2}+\nu_{3}^{2}\right)\\
* & * & -2\nu_{4}\left(c_{2}\nu_{2}+c_{1}\nu_{1}\right) & \nu_{3}\left(c_{1}\nu_{1}+c_{2}\nu_{2}\right)-c_{3}\left(\nu_{1}^{2}+\nu_{2}^{2}\right)\\
* & * & * & 0
\end{array}\right).
\end{equation}
The condition $\mathbf{h}\cdot\mathbf{p}=0$ implies $\bm{\beta}^{T}A\bm{\beta}=0$, which leads to a necessary and sufficient condition $A^{T}=-A$.
Because $A$ is symmetric and antisymmetric at the same time, it follows that all the entities of $A$ are zeros ($A_{ij}=0$).
Using expressions on the main diagonal:
\begin{eqnarray}
    c_{1}\nu_{1}+c_{2}\nu_{2}=0, \quad c_{1}\nu_{1}+c_{3}\nu_{3}=0, \quad c_{2}\nu_{2}+c_{3}\nu_{3} = 0,
\end{eqnarray}
and substituting them into the fourth column of $A$, we get
\begin{eqnarray}
    c_{1}\left(\nu_{2}^{2}+\nu_{3}^{2}\right)=0, \quad c_{2}\left(\nu_{1}^{2}+\nu_{3}^{2}\right)=0, \quad c_{3}\left(\nu_{1}^{2}+\nu_{2}^{2}\right)=0,
\end{eqnarray}
that for arbitrary vector $\bm{\nu}$ allows us to deduce $c_{1}=c_{2}=c_{3}=0$.
Therefore, it is impossible to construct a vector $\mathbf{h}$ orthogonal to $\bm{\nu}$, $\mathbf{p}$, $\mathbf{d}$ which is linear in terms of components of $\bm{\beta}$.

\section{LLG equation for angle variables}\label{A:LLG_angles}
Employing  the magnetization parametrization with angles $(\Theta,\Phi,\Psi)$, we can obtain the following energy variations:
\begin{eqnarray}
     &&\dfrac{1}{\mathcal{M}}\dfrac{\delta\mathcal{E}}{\delta\Theta}=\dfrac{1}{\mathcal{M}}\dfrac{\delta\mathcal{E}}{\delta \nu_{1}}\dfrac{\partial \nu_{1}}{\partial\Theta}+\dfrac{1}{\mathcal{M}}\dfrac{\delta\mathcal{E}}{\delta \nu_{2}}\dfrac{\partial \nu_{2}}{\partial\Theta}+\dfrac{1}{\mathcal{M}}\dfrac{\delta\mathcal{E}}{\delta \nu_{3}}\dfrac{\partial \nu_{3}}{\partial\Theta}=-\left(\beta_{1}\cos\Phi+\beta_{2}\sin\Phi\right)\cos\Theta\cos\Psi+\beta_{3}\sin\Theta\cos\Psi,\nonumber\\
     &&\dfrac{1}{\mathcal{M}}\dfrac{\delta\mathcal{E}}{\delta\Phi}=\dfrac{1}{\mathcal{M}}\dfrac{\delta\mathcal{E}}{\delta \nu_{1}}\dfrac{\partial \nu_{1}}{\partial\Phi}+\dfrac{1}{\mathcal{M}}\dfrac{\delta\mathcal{E}}{\delta \nu_{2}}\dfrac{\partial \nu_{2}}{\partial\Phi}=\left(\beta_{1}\sin\Phi-\beta_{2}\cos\Phi\right)\sin\Theta\cos\Psi,\\
     &&\dfrac{1}{\mathcal{M}}\dfrac{\delta\mathcal{E}}{\delta\Psi}=\dfrac{1}{\mathcal{M}}\dfrac{\delta\mathcal{E}}{\delta \nu_{1}}\dfrac{\partial \nu_{1}}{\partial\Psi}+\dfrac{1}{\mathcal{M}}\dfrac{\delta\mathcal{E}}{\delta \nu_{2}}\dfrac{\partial \nu_{2}}{\partial\Psi}+\dfrac{1}{\mathcal{M}}\dfrac{\delta\mathcal{E}}{\delta \nu_{3}}\dfrac{\partial \nu_{3}}{\partial\Psi}+\dfrac{1}{\mathcal{M}}\dfrac{\delta\mathcal{E}}{\delta \nu_{4}}\dfrac{\partial \nu_{4}}{\partial\Psi}\nonumber\\
     &&=\left(\beta_{1}\sin\Theta\cos\Phi+\beta_{2}\sin\Theta\sin\Phi+\beta_{3}\cos\Theta\right)\sin\Psi-\beta_{4}\cos\Psi.\nonumber\label{varE_TFP}
\end{eqnarray}
Spatial and time derivatives of vector $\bm{\nu}$ are given by:
\begin{eqnarray}
    && \partial \nu_{1} = \left(\partial\Theta\cos\Theta\cos\Phi-\partial\Phi\sin\Theta\sin\Phi\right)\cos\Psi-\partial\Psi\sin\Theta\cos\Phi\sin\Psi,\nonumber\\
    && \partial \nu_{2} = \left(\partial\Theta\cos\Theta\sin\Phi+\partial\Phi\sin\Theta\cos\Phi\right)\cos\Psi-\partial\Psi\sin\Theta\sin\Phi\sin\Psi,\nonumber\\
    && \partial \nu_{3}=-\partial\Theta\sin\Theta\cos\Psi-\partial\Psi\cos\Theta\sin\Psi,\nonumber\\
    && \partial \nu_{4}=\partial\Psi \cos\Psi.\label{derivative_TFP}
\end{eqnarray}
The cross-product terms appearing in the Zhang-Li torque Eq.~(11) in the main text are given by:
\begin{eqnarray}
    \left(\begin{array}{c}
\nu_{2}\partial \nu_{3}-\nu_{3}\partial \nu_{2}\\
\nu_{3}\partial \nu_{1}-\nu_{1}\partial \nu_{3}\\
\nu_{1}\partial \nu_{2}-\nu_{2}\partial \nu_{1}
\end{array}\right)=\left(\begin{array}{c}
-\partial\Theta\sin\Phi-\partial\Phi\sin\Theta\cos\Theta\cos\Phi\\
\partial\Theta\cos\Phi-\partial\Phi\sin\Theta\cos\Theta\sin\Phi\\
\partial\Phi\sin^{2}\Theta
\end{array}\right)\cos^{2}\Psi.\label{m_time_nabla_m_TFP}
\end{eqnarray}
Now \eqref{derivative_TFP} and \eqref{m_time_nabla_m_TFP} can be plugged into Eqs.~(9), (11) in the main text.
We immediately obtain the equation for $\dot{\Psi}$:
\begin{eqnarray}
    \dot{\Psi}=-\dfrac{\alpha\gamma}{\mathcal{M}}\dfrac{\delta\mathcal{E}}{\delta\Psi} + \dfrac{1+\xi\alpha}{1+\alpha^{2}}(\mathbf{u}\cdot\nabla)\Psi.\label{RLLG_Psi}
\end{eqnarray}
To obtain equations for $\dot{\Theta}$ and $\dot{\Phi}$, we consider different combinations of Eq.~(9) in the main text:
\begin{eqnarray}
    &&-\dot{{\nu}}_{1}\sin\Phi+\dot{{\nu}}_{2}\cos\Phi=\dot{\Phi}\sin\Theta\cos\Psi\nonumber\\
    &&=\dfrac{\gamma}{\mathcal{M}}\dfrac{\delta\mathcal{E}}{\delta\Theta}-\dfrac{\alpha\gamma}{\mathcal{M}\sin\Theta\cos\Psi}\dfrac{\delta\mathcal{E}}{\delta\Phi}-\dfrac{\xi-\alpha}{1+\alpha^{2}}\cos^{2}\Psi(\mathbf{u}\cdot\nabla)\Theta+\dfrac{1+\xi\alpha}{1+\alpha^{2}}\sin\Theta\cos\Psi(\mathbf{u}\cdot\nabla)\Phi.\label{Phi_1}
\end{eqnarray}
And also:
\begin{eqnarray}
    &&\dot{{\nu}}_{1}\cos\Phi+\dot{{\nu}}_{2}\sin\Phi=\dot{\Theta}\cos\Theta\cos\Psi-\dot{\Psi}\sin\Theta\sin\Psi\nonumber\\
    &&=-\dfrac{\gamma}{\mathcal{M}}\dfrac{\cos\Theta}{\sin\Theta}\dfrac{\delta\mathcal{E}}{\delta\Phi}-\alpha\gamma\left(\sin\Theta\cos\Psi\left(\bm{\nu}\cdot\bm{\beta}\right)-\beta_{1}\cos\Phi-\beta_{2}\sin\Phi\right)\nonumber\\
    &&
    +\dfrac{\xi-\alpha}{1+\alpha^{2}}\sin\Theta\cos\Theta\cos^{2}\Psi(\mathbf{u}\cdot\nabla)\Phi + \dfrac{1+\xi\alpha}{1+\alpha^{2}}\left[\cos\Theta\cos\Psi(\mathbf{u}\cdot\nabla)\Theta-\sin\Theta\sin\Psi(\mathbf{u}\cdot\nabla)\Psi\right].\label{ThetaPsi_1}
\end{eqnarray}
The equation for $\dot{{\nu}}_{3}$ in Eq.~(9) in the main text has the form:
\begin{eqnarray}
 && \dot{n}_\mathrm{z} =-\dot{\Theta}\sin\Theta\cos\Psi-\dot{\Psi}\cos\Theta\sin\Psi \nonumber\\
 && = \dfrac{\gamma}{\mathcal{M}}\dfrac{\delta\mathcal{E}}{\delta\Phi}-\alpha\gamma\left(\cos\Theta\cos\Psi\left(\bm{\nu}\cdot\bm{\beta}\right)-\beta_{3}\right)\nonumber\\
 &&-\dfrac{\xi-\alpha}{1+\alpha^{2}}\sin^{2}\Theta\cos^{2}\Psi(\mathbf{u}\cdot\nabla)\Phi -\dfrac{1+\xi\alpha}{1+\alpha^{2}}\left(\sin\Theta\cos\Psi(\mathbf{u}\cdot\nabla)\Theta+\cos\Theta\sin\Psi(\mathbf{u}\cdot\nabla)\Psi\right).\label{ThetaPsi_2}
\end{eqnarray}
Then multiplying \eqref{ThetaPsi_1} and \eqref{ThetaPsi_2} by $\cos\Theta$ and $-\sin\Theta$, respectively, and adding them, we can get:
\begin{eqnarray}
    &&\dot{\Theta}\sin\Theta\cos\Psi=-\dfrac{\gamma}{\mathcal{M}}\dfrac{\delta\mathcal{E}}{\delta\Phi}-\dfrac{\alpha\gamma}{\mathcal{M}}\dfrac{\sin\Theta}{\cos\Psi}\dfrac{\delta\mathcal{E}}{\delta\Theta}+\dfrac{\xi-\alpha}{1+\alpha^{2}}\sin^{2}\Theta\cos^{2}\Psi(\mathbf{u}\cdot\nabla)\Phi+\dfrac{1+\xi\alpha}{1+\alpha^{2}}\sin\Theta\cos\Psi(\mathbf{u}\cdot\nabla)\Theta.\label{Theta_1}
\end{eqnarray}
As we can see, Eqs.~\eqref{Phi_1}, \eqref{Theta_1} already represent dynamic equations for $\Phi$ and $\Theta$ variables. 
To write these equations in a canonical way, we have to exclude variational derivatives, $\delta\mathcal{E}/\delta\Theta$ from \eqref{Phi_1} and $\delta\mathcal{E}/\delta\Phi$ from \eqref{Theta_1}. 
So, from \eqref{Phi_1} and \eqref{Theta_1}, we find:
\begin{eqnarray}
    && \dfrac{\gamma}{\mathcal{M}}\dfrac{\delta\mathcal{E}}{\delta\Theta}=\dot{\Phi}\sin\Theta\cos\Psi+\dfrac{\alpha\gamma}{\mathcal{M}\sin\Theta\cos\Psi}\dfrac{\delta\mathcal{E}}{\delta\Phi}+\dfrac{\xi-\alpha}{1+\alpha^{2}}\cos^{2}\Psi(\mathbf{u}\cdot\nabla)\Theta-\dfrac{1+\xi\alpha}{1+\alpha^{2}}\sin\Theta\cos\Psi(\mathbf{u}\cdot\nabla)\Phi,\nonumber\\
    && \dfrac{\gamma}{\mathcal{M}}\dfrac{\delta\mathcal{E}}{\delta\Phi}=-\dot{\Theta}\sin\Theta\cos\Psi-\dfrac{\alpha\gamma\sin\Theta}{\mathcal{M}\cos\Psi}\dfrac{\delta\mathcal{E}}{\delta\Theta}+\dfrac{\xi-\alpha}{1+\alpha^{2}}\sin^{2}\Theta\cos^{2}\Psi(\mathbf{u}\cdot\nabla)\Phi+\dfrac{1+\xi\alpha}{1+\alpha^{2}}\sin\Theta\cos\Psi(\mathbf{u}\cdot\nabla)\Theta,
\end{eqnarray}
that can be plugged into \eqref{Theta_1} and \eqref{Phi_1}:
\begin{eqnarray}
    &&\dot{\Theta}\sin\Theta\cos\Psi=-\dfrac{\gamma}{\mathcal{M}}\left(1+\dfrac{\alpha^{2}}{\cos^{2}\Psi}\right)\dfrac{\delta\mathcal{E}}{\delta\Phi}-\alpha\dot{\Phi}\sin^{2}\Theta+\sin\Theta\cos\Psi(\mathbf{u}\cdot\nabla)\Theta+F(\Psi)\sin^{2}\Theta(\mathbf{u}\cdot\nabla)\Phi,\nonumber\\
    &&\dot{\Phi}\sin\Theta\cos\Psi=\dfrac{\gamma}{\mathcal{M}}\left(1+\dfrac{\alpha^{2}}{\cos^{2}\Psi}\right)\dfrac{\delta\mathcal{E}}{\delta\Theta}+\alpha\dot{\Theta}+\sin\Theta\cos\Psi(\mathbf{u}\cdot\nabla)\Phi-F(\Psi)(\mathbf{u}\cdot\nabla)\Theta,\label{TF_eqs}
\end{eqnarray}
where $F(\Psi)=\left[\alpha+\xi\alpha^{2}+(\xi-\alpha)\cos^{2}\Psi\right]/\left(1+\alpha^{2}\right)$.
The connection to the standard LLG equation Eq.~(2) in the main text with the Zhang-Li torque Eq.~(10) in the main text can be easily seen from \eqref{TF_eqs} letting $\Psi=0$ providing $F=\xi$:
\begin{eqnarray}
    &&\dot{\Theta}\sin\Theta=-\dfrac{\gamma\left(1+\alpha^{2}\right)}{\mathcal{M}}\dfrac{\delta E}{\delta\Phi}-\alpha\dot{\Phi}\sin^{2}\Theta+\sin\Theta(\mathbf{u}\cdot\nabla)\Theta+\xi\sin^{2}\Theta(\mathbf{u}\cdot\nabla)\Phi,\nonumber\\
    &&\dot{\Phi}\sin\Theta=\dfrac{\gamma\left(1+\alpha^{2}\right)}{\mathcal{M}}\dfrac{\delta E}{\delta\Theta}+\alpha\dot{\Theta}+\sin\Theta(\mathbf{u}\cdot\nabla)\Phi-\xi(\mathbf{u}\cdot\nabla)\Theta.\label{TF_eqs_standard}
\end{eqnarray}
The system of equations \eqref{TF_eqs_standard} represents the classical LLG equation with the Zhang-Li torque written in Slonczewski form.

\section{Generalized Thiele equation}\label{A:Thiele}
We express the variational derivatives of the Hamiltonian from the regularized LLG equation~\eqref{TF_eqs}, \eqref{RLLG_Psi}:
\begin{eqnarray}
    && \dfrac{\alpha\gamma}{\mathcal{M}}\dfrac{\delta\mathcal{E}}{\delta\Phi}=\dfrac{\alpha\cos^{2}\Psi}{\cos^{2}\Psi+\alpha^{2}}\left[-\dot{\Theta}\sin\Theta\cos\Psi-\alpha\dot{\Phi}\sin^{2}\Theta+\mathcal{I}_{\Theta}\right],\nonumber\\
    &&\dfrac{\alpha\gamma}{\mathcal{M}}\dfrac{\delta\mathcal{E}}{\delta\Theta}=\dfrac{\alpha\cos^{2}\Psi}{\cos^{2}\Psi+\alpha^{2}}\left[\dot{\Phi}\sin\Theta\cos\Psi-\alpha\dot{\Theta}-\mathcal{I}_{\Phi}\right], \nonumber\\
    && \dfrac{\alpha\gamma}{\mathcal{M}}\dfrac{\delta\mathcal{E}}{\delta\Psi}=-\dot{\Psi} + \dfrac{1+\xi\alpha}{1+\alpha^{2}}(\mathbf{u}\cdot\nabla)\Psi,
    \\
    && \mathcal{I}_{\Theta}=\sin\Theta\cos\Psi(\mathbf{u}\cdot\nabla)\Theta+F(\Psi)\sin^{2}\Theta(\mathbf{u}\cdot\nabla)\Phi,\nonumber\\
    &&\mathcal{I}_{\Phi}=\sin\Theta\cos\Psi(\mathbf{u}\cdot\nabla)\Phi-F(\Psi)(\mathbf{u}\cdot\nabla)\Theta.\nonumber
\end{eqnarray}
Then we plug this into the dissipation integral Eq.~(14) in the main text with taking into account the assumption about the rigid motion Eq.~(13) in the main text:
\begin{eqnarray}
    && \dfrac{\alpha\gamma}{\mathcal{M}}\left(\dfrac{\delta\mathcal{E}}{\delta\Theta}\nabla\Theta+\dfrac{\delta\mathcal{E}}{\delta\Phi}\nabla\Phi+\dfrac{\delta\mathcal{E}}{\delta\Psi}\nabla\Psi\right) = \dfrac{\alpha\cos^{2}\Psi}{\cos^{2}\Psi+\alpha^{2}}\left[\mathbf{v}\cdot\nabla\Theta\sin\Theta\cos\Psi+\alpha\mathbf{v}\cdot\nabla\Phi\sin^{2}\Theta+\mathcal{I}_{\Theta}\right]\nabla\Phi\nonumber\\
    &&+\dfrac{\alpha\cos^{2}\Psi}{\cos^{2}\Psi+\alpha^{2}}\left[-\mathbf{v}\cdot\nabla\Phi\sin\Theta\cos\Psi+\alpha\mathbf{v}\cdot\nabla\Theta-\mathcal{I}_{\Phi}\right]\nabla\Theta+\left[\mathbf{v}\cdot\nabla\Psi+ \dfrac{1+\xi\alpha}{1+\alpha^{2}}(\mathbf{u}\cdot\nabla)\Psi\right]\nabla\Psi.\label{Force_TFP}
\end{eqnarray}
We can simplify these expressions by noticing the following:
\begin{eqnarray}
    &&(\mathbf{v}\cdot\nabla\Theta)\nabla\Phi-(\mathbf{v}\cdot\nabla\Phi)\nabla\Theta=\mathbf{v}\times\left(\nabla\Phi\times\nabla\Theta\right),\nonumber\\
    &&\epsilon_{ijk}\mathbf{n}\cdot\partial_{j}\mathbf{n}\times\partial_{k}\mathbf{n}=-\sin\Theta\cos^{3}\Psi\left(\nabla\Phi\times\nabla\Theta\right)_{i},\nonumber\\
    && \partial_{i}\mathbf{n}\cdot\partial_{j}\mathbf{n}=\left(\partial_{i}\Theta\partial_{j}\Theta+\partial_{i}\Phi\partial_{j}\Phi\sin^{2}\Theta\right)\cos^{2}\Psi + \partial_{i}\Psi\partial_{j}\Psi\sin^{2}\Psi,\nonumber\\
    &&\partial_{i}\nu_{4}\partial_{j}\nu_{4}=\partial_{i}\Psi\partial_{j}\Psi\cos^{2}\Psi=n^{2}\partial_{i}\Psi\partial_{j}\Psi,
\end{eqnarray}
and one part of \eqref{Force_TFP} can be integrated as:
\begin{eqnarray}
    &&\intop\dfrac{\cos^{3}\Psi}{\cos^{2}\Psi+\alpha^{2}}\left(\mathbf{v}\times\left(\nabla\Phi\times\nabla\Theta\right)\sin\Theta+\mathbf{u}\times\left(\nabla\Phi\times\nabla\Theta\right)\sin\Theta\right)\mathrm{d}V=\mathbf{g}\times\left[\mathbf{v}+\mathbf{u}\right],
\end{eqnarray}
where generalized gyro-vector, $\mathbf{g}$, has components:
\begin{equation}
    g_{i}=\intop\dfrac{\sin\Theta\cos^{3}\Psi}{\cos^{2}\Psi+\alpha^{2}}\left(\nabla\Theta\times\nabla\Phi\right)_{i}\mathrm{d}V=\epsilon_{ijk}\intop\dfrac{\mathbf{n}\cdot\partial_{j}\mathbf{n}\times\partial_{k}\mathbf{n}}{n^{2}+\alpha^{2}}\mathrm{d}V, \,\, \{i,j,k\}\in\{x,y,z\}.\label{gyro_def}
\end{equation}
Then we can simplify:
\begin{eqnarray}
    &&\dfrac{\alpha^{2}\cos^{2}\Psi}{\cos^{2}\Psi+\alpha^{2}}\left(\partial_{i}\Theta\partial_{j}\Theta+\partial_{i}\Phi\partial_{j}\Phi\sin^{2}\Theta\right)v_{j}+\partial_{i}\Psi\partial_{j}\Psi  v_{j}=\dfrac{\alpha^{2} \partial_{i}\bm{\nu}\cdot\partial_{j}\bm{\nu}+\partial_{i}\nu_{4}\partial_{j}\nu_{4}}{n^{2}+\alpha^2}v_{j},\label{dis_1}
\end{eqnarray}
and the electric current part
\begin{eqnarray}
    &&\dfrac{\alpha F(\Psi)\cos^{2}\Psi}{\cos^{2}\Psi+\alpha^2}\left(\partial_{j}\Theta\partial_{k}\Theta+\partial_{j}\Phi\partial_{k}\Phi\sin^{2}\Theta\right)u_{k}+\dfrac{1+\xi\alpha}{1+\alpha^{2}}\partial_{j}\Psi\partial_{k}\Psi\cdot u_{k}=\nonumber\\
    && \dfrac{1+\xi\alpha}{1+\alpha^{2}}\dfrac{\alpha^{2} \partial_{j}\bm{\nu}\cdot\partial_{k}\bm{\nu}+\partial_{j}\nu_{4}\partial_{k}\nu_{4}}{n^{2}+\alpha^2}u_{k}+\dfrac{\alpha(\xi-\alpha)}{1+\alpha^{2}}\dfrac{n^{2}\partial_{j}\bm{\nu}\cdot\partial_{k}\bm{\nu}-\partial_{j}\nu_{4}\partial_{k}\nu_{4}}{n^{2}+\alpha^{2}}u_{k}.\label{dis_2}
\end{eqnarray}
Then, using \eqref{gyro_def}, \eqref{dis_1} and \eqref{dis_2}, we can integrate \eqref{Force_TFP} and present it in the vector form:
\begin{equation}
    \alpha\mathbf{g}\times\left[\mathbf{v}+\mathbf{u}\right]+\hat{\gamma}\mathbf{v}+\left[\dfrac{1+\xi\alpha}{1+\alpha^{2}}\hat{\gamma}+\dfrac{\xi-\alpha}{1+\alpha^{2}}\hat{\gamma}^{\prime}\right]\mathbf{u}=0,\label{gen_Thiele}
\end{equation}
where generalized dissipation tensors, $\hat{\gamma}$ and $\hat{\gamma}^{\prime}$ have components:
\begin{equation}
    \gamma_{jk}=\intop\dfrac{\alpha^{2} \partial_{j}\bm{\nu}\cdot\partial_{k}\bm{\nu}+\partial_{j}\nu_{4}\partial_{k}\nu_{4}}{n^{2}+\alpha^2}\mathrm{d}V,\,\,\gamma_{jk}^{\prime}=\alpha\intop\dfrac{n^{2}\partial_{j}\bm{\nu}\cdot\partial_{k}\bm{\nu}-\partial_{j}\nu_{4}\partial_{k}\nu_{4}}{n^{2}+\alpha^2}\mathrm{d}V.
\end{equation}
To see the connection \eqref{gen_Thiele} with the standard Thiele equation, we just let $\nu_{4}=0$ and $n=1$, and get:
\begin{eqnarray}
    && g_{i}=\dfrac{\epsilon_{ijk}}{1+\alpha^{2}}\intop\mathbf{n}\cdot\partial_{j}\mathbf{n}\times\partial_{k}\mathbf{n}\mathrm{d}V=\dfrac{G_{i}}{1+\alpha^{2}},\,\,\hat{\gamma}=\dfrac{\alpha^{2}}{1+\alpha^{2}}\Gamma,\,\,\hat{\gamma}^{\prime}=\dfrac{\alpha}{1+\alpha^{2}}\Gamma,\,\,\Gamma_{jk}=\intop\partial_{j}\mathbf{n}\cdot\partial_{k}\mathbf{n}\,\mathrm{d}V,
\end{eqnarray}
where $\Gamma$ is a standard dissipation tensor with components, $\Gamma_{jk}$.
So the Thiele equation is written as:
\begin{equation}
    \mathbf{G}\times\left[\mathbf{v}+\mathbf{u}\right]+\Gamma\left[\alpha\mathbf{v}+\xi\mathbf{u}\right]=0.
\end{equation}

\section{Solutions of the Thiele equation for the in- and out-of-plane motions}\label{A:Thiele_sol}
In the case of in-plane current $\mathbf{u}=u\mathbf{e}_\mathrm{x}$, the Thiele equation Eq.~(15) in the main text can be written in the matrix form as:
\begin{eqnarray}
    \left[\hat{\gamma}+\alpha\left(\begin{array}{ccc}
0 & -g_{z} & g_{{y}}\\
g_{z} & 0 & -g_{{x}}\\
-g_{{y}} & g_{\mathrm{x}} & 0
\end{array}\right)\right]\left(\begin{array}{c}
v_{\mathrm{x}}\\
v_{\mathrm{y}}\\
v_{\mathrm{z}}
\end{array}\right)+\left[\xi\hat{\gamma}^{\prime\prime}+\alpha\left(\begin{array}{ccc}
0 & -g_{z} & g_{{y}}\\
g_{z} & 0 & -g_{{x}}\\
-g_{{y}} & g_{{x}} & 0
\end{array}\right)\right]\left(\begin{array}{c}
u\\
0\\
0
\end{array}\right)=0,\label{Thiele_matrix1}
\end{eqnarray}
where we introduced tensor $\hat{\gamma}^{\prime}=\left[\dfrac{1+\xi\alpha}{1+\alpha^{2}}\hat{\gamma}+\dfrac{\xi-\alpha}{1+\alpha^{2}}\hat{\gamma}^{\prime}\right]/\xi$.
Now, we consider a special case when the soliton moves in the plane only, i.e., $\mathbf{v}=(v_\mathrm{x},v_\mathrm{y},0)$.
In this case, \eqref{Thiele_matrix1} allows to get:
\begin{equation}
    \left(\begin{array}{cc}
\gamma_{\mathrm{xx}} & \gamma_{\mathrm{xy}}-\alpha g_\mathrm{z}\\
\gamma_{\mathrm{xy}}+\alpha g_{z} & \gamma_{\mathrm{yy}}
\end{array}\right)\left(\begin{array}{c}
v_{\mathrm{x}}\\
v_{\mathrm{y}}
\end{array}\right)=-u\left(\begin{array}{c}
\xi\gamma_{\mathrm{xx}}^{\prime\prime}\\
\xi\gamma_{\mathrm{xy}}^{\prime\prime}+\alpha g_{\mathrm{z}}
\end{array}\right),\label{Thiele_matrix3}
\end{equation}
that can be solved as:
\begin{eqnarray}
    &&\left(\begin{array}{c}
v_{\mathrm{x}}\\
v_{\mathrm{y}}
\end{array}\right)=\dfrac{-u}{\Delta+\alpha^{2}g_{\mathrm{z}}^{2}}\left(\begin{array}{cc}
\gamma_{\mathrm{yy}} & -\gamma_{\mathrm{xy}}+\alpha g_{\mathrm{z}}\\
-\gamma_{\mathrm{xy}}-\alpha g_{\mathrm{z}} & \gamma_{\mathrm{xx}}
\end{array}\right)\left(\begin{array}{c}
\xi\gamma_{\mathrm{xx}}^{\prime\prime}\\
\xi\gamma_{\mathrm{xy}}^{\prime\prime}+\alpha g_{\mathrm{z}}
\end{array}\right)\nonumber\\&&=\dfrac{-u}{\Delta+\alpha^{2}g_{{z}}^{2}}\left(\begin{array}{c}
\xi\left(\gamma_{\mathrm{yy}}\gamma_{\mathrm{xx}}^{\prime\prime}-\gamma_{\mathrm{xy}}\gamma_{\mathrm{xy}}^{\prime\prime}\right)-\alpha\left(\gamma_{\mathrm{xy}}-\xi\gamma_{\mathrm{xy}}^{\prime\prime}\right)g_{{z}}+\alpha^{2}g_{{z}}^{2}\\
\xi\left(\gamma_{\mathrm{xx}}\gamma_{\mathrm{xy}}^{\prime\prime}-\gamma_{\mathrm{xx}}^{\prime\prime}\gamma_{\mathrm{xy}}\right)+\alpha\left(\gamma_{\mathrm{xx}}-\xi\gamma_{\mathrm{xx}}^{\prime\prime}\right)g_{\mathrm{z}}
\end{array}\right),
\end{eqnarray}
where $\Delta=\gamma_{\mathrm{xx}}\gamma_{\mathrm{yy}}-\left(\gamma_{\mathrm{xy}}\right)^{2}$.
This solution simplifies in the case $\xi=\alpha$, then one has $\gamma^{\prime\prime}=\gamma/\xi$, and velocity is $\mathbf{v}=(-u,0,0)$.

In a general case, $\mathbf{v}=(v_\mathrm{x},v_\mathrm{y}, v_\mathrm{z})$, and solution of \eqref{Thiele_matrix1} can be written as:
\begin{eqnarray}
    &\left(\begin{array}{c}
v_{\mathrm{x}}\\
v_{\mathrm{y}}\\
v_{\mathrm{z}}
\end{array}\right)=-\dfrac{u}{\Omega}A\left(\begin{array}{c}
\xi\gamma_{xx}^{\prime\prime}\\
\xi\gamma_{xy}^{\prime\prime}+\alpha g_{z}\\
\xi\gamma_{xz}^{\prime\prime}-\alpha g_{y}
\end{array}\right)\\
&-\dfrac{u\alpha}{\Omega}\left(\begin{array}{c}
\left(g_{x}\gamma_{xz}+g_{y}\gamma_{yz}+g_{z}\gamma_{zz}\right)\left(\xi\gamma_{xy}^{\prime\prime}+\alpha g_{z}\right)-\left(g_{x}\gamma_{xy}+g_{y}\gamma_{yy}+g_{z}\gamma_{yz}\right)\left(\xi\gamma_{xz}^{\prime\prime}-\alpha g_{y}\right)\\
\left(g_{x}\gamma_{xx}+g_{y}\gamma_{xy}+g_{z}\gamma_{xz}\right)\left(\xi\gamma_{xz}^{\prime\prime}-\alpha g_{y}\right)-\xi\gamma_{xx}^{\prime\prime}\left(g_{x}\gamma_{xz}+g_{y}\gamma_{yz}+g_{z}\gamma_{zz}\right)\\
\xi\gamma_{xx}^{\prime\prime}\left(g_{x}\gamma_{xy}+g_{y}\gamma_{yy}+g_{z}\gamma_{yz}\right)-\left(g_{x}\gamma_{xx}+g_{y}\gamma_{xy}+g_{z}\gamma_{xz}\right)\left(\xi\gamma_{xy}^{\prime\prime}+\alpha g_{z}\right)
\end{array}\right),\nonumber
\end{eqnarray}
where matrix $A$ has entities $A_{ii}=\gamma_{jj}\gamma_{kk}-\gamma_{jk}^{2}+\alpha^{2}g_{i}^{2}$ and $A_{ij}=\gamma_{ik}\gamma_{jk}-\gamma_{kk}\gamma_{ij}+\alpha^{2}g_{i}g_{j}, i\neq j\neq k$, and $\Omega$ is given by expression:
\begin{eqnarray}
&\Omega=\gamma_{xx}\gamma_{yy}\gamma_{zz}-\gamma_{xx}\left(\gamma_{yz}^{2}-\alpha^{2}g_{x}^{2}\right)-\gamma_{yy}\left(\gamma_{xz}^{2}-\alpha^{2}g_{y}^{2}\right)-\gamma_{zz}^{\prime}\left(\gamma_{xy}^{2}-\alpha^{2}g_{z}^{2}\right)\label{Omega}\\&+\left(\gamma_{yz}-\alpha g_{x}\right)\left(\gamma_{xz}-\alpha g_{y}\right)\left(\gamma_{xy}-\alpha g_{z}\right)+\left(\gamma_{yz}+\alpha g_{x}\right)\left(\gamma_{xz}+\alpha g_{y}\right)\left(\gamma_{xy}+\alpha g_{z}\right).\nonumber
\end{eqnarray}

\begin{center}
\vspace{0.5em}
\textbf{Supplementary Table 1: Comparison of numerical simulations and Thiele approach results}
\begin{tabular}{|c|c|c|c|c|c|c|}
\hline
\multirow{2}{*}{$c_{i}$} & \multicolumn{2}{c|}{\textbf{Skyrmion tube}} & \multicolumn{2}{c|}{\textbf{Chiral bobber}} & \multicolumn{2}{c|}{\textbf{Dipole string}} \\
\cline{2-7}
 & Numerics & Thiele approach & Numerics & Thiele approach & Numerics & Thiele approach \\
\hline
$c_{x}$ & 1.0018 & 1.0058 & 1.1478 & 1.1434  & 1.1418 & 1.1254 \\
\hline
$c_{y}$ & 0.2666 & 0.2657 & 0.1618 & 0.1419  & 0.1112 & 0.07559 \\
\hline
$c_{z}$ & 0 & 0 & 0 & 0 & -0.03385 & -0.02247 \\
\hline
\end{tabular}
\end{center}

\begin{figure*}[tb!]
    \centering
    \includegraphics[width=17.8cm]{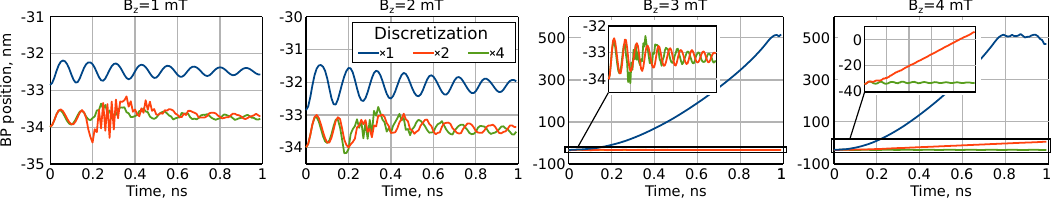}
\caption{ ~\small \textbf{BP dynamics in a nanowire}. Results of the dynamics of a BP in a Fe nanowire are shown. 
% %
The dynamics governed by the classical LLG equation is induced by external magnetic fields $B_\mathrm{z}$ of strengths $1,2,3$ and $4$ mT. 
The position of the BP is calculated from Eq.~(21) in the main text.
Simulations were conducted with different levels of discretization: 
$\times1$ corresponds to a grid of 
$16\times 16\times512$, while 
$\times2$ and $\times4$ represent scalings of each dimension by factors of 2 and 4, respectively.}
\label{FigBP}
\end{figure*}

\begin{figure*}[tb!]
\centering

\end{figure*}

\end{document}